\documentclass[twocolumn]{aastex63}
\submitjournal{ApJ}

\usepackage{hhline}
\usepackage{hyperref}
\usepackage{amsmath}
\usepackage{amssymb}
\usepackage{graphicx}
\usepackage{xspace}

\newcommand{\Teff}{T_{\rm eff}}
\newcommand{\logg}{\log g}
\newcommand{\unit}[1]{\ensuremath{\mathrm{\,#1}}\xspace}
\newcommand{\kms}{\unit{km\,s^{-1}}}
\newcommand{\kmsec}{\kms}

\newcommand{\Gaia}{\textit{Gaia}\xspace}

\newcommand{\tmix}{\ensuremath{\tau_{\rm mix}}\xspace}
\newcommand{\tsf}{\ensuremath{\tau_{\rm sf}}\xspace}

\newcommand{\unifdistr}[2]{\mathcal{U}\left[#1,#2\right]}

\defcitealias{Simon15}{S15}
\defcitealias{Koposov15b}{K15}
\defcitealias{Walker15}{W15}
\defcitealias{Ji16c}{J16}
\defcitealias{Roederer16b}{R16}

\shorttitle{Metal Mixing in the UFD Reticulum II}
\shortauthors{Ji et al.}

\begin{document}
\title{Metal Mixing in the R-Process Enhanced Ultra-Faint Dwarf Galaxy Reticulum II\footnote{
This paper includes data gathered with the 6.5~meter 
Magellan Telescopes located at Las Campanas Observatory, Chile.}}

\correspondingauthor{Alexander P. Ji}
\email{alexji@uchicago.edu}

\author[0000-0002-4863-8842]{Alexander P. Ji}
\affiliation{Department of Astronomy \& Astrophysics, University of Chicago, 5640 S Ellis Avenue, Chicago, IL 60637, USA}
\affiliation{Kavli Institute for Cosmological Physics, University of Chicago, Chicago, IL 60637, USA}
\affiliation{Joint Institute for Nuclear Astrophysics -- Center for the Evolution of the Elements (JINA-CEE), USA}
\author[0000-0002-4733-4994]{Joshua D. Simon}
\affiliation{Observatories of the Carnegie Institution for Science, 813 Santa Barbara St., Pasadena, CA 91101, USA}
\author[0000-0001-5107-8930]{Ian U. Roederer}
\affiliation{Department of Astronomy, University of Michigan, 1085 S. University Ave., Ann Arbor, MI 48109, USA}
\affiliation{Joint Institute for Nuclear Astrophysics -- Center for the Evolution of the Elements (JINA-CEE), USA}
\author[0000-0001-8784-6794]{Ekaterina Magg}
\affiliation{Max Planck Institute for Astronomy, K\"{o}nigstuhl 17, Heidelberg, D-69117, Germany}
\author[0000-0002-2139-7145]{Anna Frebel}
\affiliation{Department of Physics \& Kavli Institute for Astrophysics and Space Research, Massachusetts Institute of Technology, Cambridge, MA 02139, USA}
\affiliation{Joint Institute for Nuclear Astrophysics -- Center for the Evolution of the Elements (JINA-CEE), USA}
\author[0000-0002-8878-3315]{Christian I. Johnson}
\affiliation{Space Telescope Science Institute, 3700 San Martin Drive, Baltimore, MD 21218, USA}
\author[0000-0002-0560-3172]{Ralf S.\ Klessen}
\affiliation{Universit\"{a}t Heidelberg, Zentrum f\"{u}r Astronomie, Institut f\"{u}r Theoretische Astrophysik, D-69120 Heidelberg, Germany}
\affiliation{Universit\"{a}t Heidelberg, Interdiszipli\"{a}res Zentrum f\"{u}r Wissenschaftliches Rechnen, D-69120 Heidelberg, Germany}
\author[0000-0002-9022-5136]{Mattis Magg}
\affiliation{Universit\"{a}t Heidelberg, Zentrum f\"{u}r Astronomie, Institut f\"{u}r Theoretische Astrophysik, D-69120 Heidelberg, Germany}
\affiliation{International Max Planck Research School for Astronomy and Cosmic Physics at the University of Heidelberg (IMPRS-HD)}
\author[0000-0002-3184-9918]{Gabriele Cescutti}
\affiliation{ Dipartimento di Fisica, Sezione di Astronomia, Università di Trieste, Via G. B. Tiepolo 11, 34143 Trieste, Italy}
\affiliation{INAF, Osservatorio Astronomico di Trieste, Via Tiepolo 11, I-34143 Trieste, Italy}
\affiliation{INFN, Sezione di Trieste, Via A. Valerio 2, I-34127 Trieste, Italy}
\author{Mario Mateo}
\affiliation{Department of Astronomy, University of Michigan, 1085 S. University Ave., Ann Arbor, MI 48109, USA}
\author[0000-0002-9908-5571]{Maria Bergemann}
\affiliation{Max Planck Institute for Astronomy, K\"{o}nigstuhl 17, Heidelberg, D-69117, Germany}
\affiliation{Niels Bohr International Academy, Niels Bohr Institute, University of Copenhagen,
Blegdamsvej 17, DK-2100 Copenhagen, Denmark}
\author[0000-0002-4272-263X]{John I. Bailey, III}
\affiliation{Department of Physics, UCSB, Santa Barbara, CA 93016, USA}

\begin{abstract}
The ultra-faint dwarf galaxy Reticulum~II was enriched by a single rare and prolific $r$-process event.
The $r$-process content of Reticulum~II thus provides a unique opportunity to study metal mixing in a relic first galaxy.
Using multi-object high-resolution spectroscopy with VLT/GIRAFFE and Magellan/M2FS, we identify 32 clear spectroscopic member stars and measure abundances of Mg, Ca, Fe, and Ba where possible.
We find $72^{+10}_{-12}\%$ of the stars are $r$-process-enhanced, with a mean $\left\langle\mbox{[Ba/H]}\right\rangle=-1.68~\pm~0.07$ and unresolved intrinsic dispersion $\sigma_{\rm [Ba/H]} < 0.20$.
The homogeneous $r$-process abundances imply that Ret~II's metals are well-mixed by the time the $r$-enhanced stars form, which simulations have shown requires at least 100 Myr of metal mixing in between bursts of star formation to homogenize.
This is the first direct evidence of bursty star formation in an ultra-faint dwarf galaxy.
The homogeneous dilution prefers a prompt and high-yield $r$-process site, such as collapsar disk winds or prompt neutron star mergers.
We also find evidence from [Ba/H] and [Mg/Ca] that the $r$-enhanced stars in Ret~II formed in the absence of substantial pristine gas accretion, perhaps indicating that ${\approx}70\%$ of Ret~II stars formed after reionization.
\end{abstract}
\keywords{nuclear reactions, nucleosynthesis, abundances --- stars: abundances --- galaxies: dwarf --- Local Group}

\section{Introduction}

Ultra-faint dwarf galaxies (UFDs) are Milky Way satellite galaxies with luminosities $M_V > -7.7$ (stellar masses ${\lesssim}10^5 M_\odot$, \citealt{Simon19}).
UFDs appear to form all their stars in the first 1-2 billion years, before their star formation is cut off by reionization \citep{Benson02,Brown14}.
UFDs probe the extreme low-mass end of galaxy formation, where star formation is inefficient and massive stars form stochastically, resulting in intermittent feedback and incomplete sampling of nucleosynthetic sources \citep{Koch08,Koch13,Simon10,Frebel10b,Frebel15,Ji16b,Ji19a}.
UFDs are also relics of early galaxy formation, providing a unique window into the first stars and galaxies in a pre-reionization universe,
as well as a clean probe of the first metal-free Population III stars \citep[e.g.,][]{Bovill2009,Salvadori2009,Frebel12,Ji15}.
Since many halo stars with $\mbox{[Fe/H]} < -2.5$ likely form in UFD-like environments, even if they later grow into or accrete into larger systems \citep{Brauer19}, it is crucial to understand the star formation conditions for UFDs to interpret the most metal-poor stars.
To understand these early properties, the red giant branch stars in UFDs have been the subject of intense spectroscopic study.
The last 15 years have resulted in high-resolution spectra of ${\gtrsim}$100 stars across ${\sim}$20 UFDs with detailed elemental abundances (see \citealt{Frebel15,Simon19,Ji19a,Ji20a} for a description of the basic characteristics and chemical evolution trends).

Reticulum~II (Ret~II) is a UFD discovered in the Dark Energy Survey \citep{Bechtol15,Koposov15a}, located only 32 kpc away.
Initial followup spectroscopy showed that its velocity dispersion, mean metallicity, and metallicity dispersion were consistent with typical UFDs (\citealt{Simon15,Koposov15b,Walker15}, henceforth \citetalias{Simon15,Koposov15b,Walker15}).
Subsequent high-resolution spectroscopy surprisingly showed that most Ret~II stars displayed some of the highest $r$-process enhancements known (\citealt{Ji16b,Ji16c,Roederer16b}, henceforth \citetalias{Ji16c,Roederer16b}).
By comparing to other UFDs (which display unusually low neutron-capture element abundances, \citealt{Frebel15,Ji19a}), the clear conclusion is that Ret~II experienced enrichment from a single rare and prolific $r$-process event.
The source of the $r$-process elements is still debated, as it could be consistent with $r$-process nucleosynthesis in a prompt neutron star merger or rare core-collapse supernova \citep[e.g.,][]{Ji16b,Beniamini16,Safarzadeh17,Safarzadeh2019b,Ojima2018,Siegel19,Tarumi2020,Molero2021,Jeon2021,Cowan2021}.

The single $r$-process event in Ret~II provides a unique opportunity to probe \emph{metal mixing} in a UFD.
Because all the $r$-process elements (including barium and europium) were deposited in a single enrichment event, the distribution of $r$/H ratios in Ret~II stars depends \emph{only} on the overall amount of enriched gas and the and homogeneity of metal mixing into the gas.
This contrasts with elements synthesized by more common sources like supernovae or asymptotic giant branch (AGB) stars, since the frequency of element production interacts with metal mixing to produce the distribution of stellar abundances \citep[e.g.,][]{Krumholz2018,Emerick2019,Emerick2020}.
A direct constraint on metal mixing by measuring the [$r$/H] distribution could play a major role in interpreting UFD abundances and formation histories \citep[e.g.,][]{Frebel12,Ji15,Webster16,Tarumi2020}.

We thus present a detailed spectroscopic study of Ret~II chemical abundances obtained with multi-object spectroscopy using VLT/FLAMES and Magellan/M2FS.
We find 32 clear member stars and 8 more candidates, the most spectroscopically confirmed members to date in Ret~II.
About half the stars have Ba and Fe constraints, while a third have Mg and/or Ca measurements as well.
Our primary focus is measuring the distribution of [Ba/H] in these stars, which is a tracer of $r$-process enrichment in Ret~II due to the high $r$-process enhancement and negligible s-process contribution in Ret~II \citep{Ji16b}.
Since this is the largest spectroscopic sample of members yet, we also explore more general kinematics, binarity, chemical evolution, and spatial gradients.
Section~\ref{sec:obsred} presents the spectroscopic observations and data reduction.
Section~\ref{sec:analysis} describes the velocity and chemical abundance analysis methods, as well as membership determination including auxiliary information from the Dark Energy Survey (DES, \citealt{desdr1}) and \Gaia EDR3 \citep{GaiaCollaboration2016,GaiaCollaboration2021,Lindegren2021}.
Section~\ref{sec:results} gives our results for the Ret~II radial velocity distribution, chemical abundance trends, Fe and Ba distributions, and radial gradients.
Section~\ref{sec:discussion} discusses the implications of our measurements on metal mixing in dwarf galaxies, the origin of the $r$-process elements, and chemical evolution in Ret~II.
We summarize and conclude in Section~\ref{sec:conclusion}.
Multi-epoch velocities are provided in Appendix~\ref{app:rv}.
A major systematic for our Ba results is microturbulence, which is discussed extensively in Appendices~\ref{app:vtba} and \ref{app:basys}.

\section{Observations and Data Reduction}\label{sec:obsred}
We observed Reticulum~II with VLT/FLAMES in October 2017 \citep{Pasquini2002}, and with Magellan/M2FS in September 2016 at medium-resolution and November 2017 at high-resolution \citep{Mateo2012}.
Table~\ref{tab:obs} contains details about which stars were observed at which settings.
Note that all signal-to-noise ratios (SNR) quoted in this paper refer to the SNR per pixel.

\subsection{VLT/FLAMES, GIRAFFE}
The FLAMES/GIRAFFE setup on the VLT UT2 provides high-resolution spectra of ${\sim}100$ stars over a field of view of diameter 0.4 degrees.
Observations were taken in visitor mode on 26-27 October 2017 with excellent weather.
We used the HR14A setting, covering one order from 6300$-$6500{\AA} with $R\sim18000$.
Targets were selected based on our own photometry of public DES Y1 images following \citet{Koposov15a}.
We chose targets near the fiducial CMD within the single field we targeted.
The total FLAMES exposure time was 11.8h, with most exposures being 3000s but a few exposures of 2400s and 3600s at the end of the night.

Data were reduced with the standard ESO pipeline, which provides flat-corrected and wavelength calibrated 1D flux and error spectra. The 1D spectra are extracted to a common rebinned dispersion without cosmic ray rejection or sky subtraction.
For each object, we removed cosmic rays in 1D by normalizing individual exposures by their median flux, then masking pixels with ${>}5\sigma$ deviations from the combined median spectrum. Care was taken not to mask pixels associated with variable sky lines.
Sky subtraction was performed in 1D mostly following \citet{Battaglia08}. For each exposure, we constructed a master sky spectrum from ${\sim}15$ sky fibers using an inverse-variance weighted mean. The master sky flux was split into two components, a sky emission line and a continuum component. The line component was used to identify wavelength bins associated with emission lines.
Then for each object spectrum, we also split the flux into emission lines and continuum, rescaled the master sky line flux to match the object emission line flux by minimizing the L1 norm (total absolute deviation at wavelengths associated with sky emission lines), applied the same scaling factor to the sky continuum, and subtracted the rescaled master sky from the object spectrum.
Visual inspection of the sky-subtracted spectra suggests this procedure was generally effective, with no correction to the line spread function or wavelength recalibration needed.
Still, there are sometimes sky subtraction residuals from spatially variable sky lines, which does impact our Ba line of interest (see Section~\ref{sec:skysub}).
Final coadded spectra were obtained using an inverse-variance weighted average of individual exposures.

\subsection{Magellan/M2FS HiRes}
We obtained high-resolution spectra of Ret~II stars with M2FS on 16-17 November 2017.
We used the HiRes mode with $180{\mu}$m slits, providing $R \sim 18000$. The detectors were binned 2x2 with 4 amplifier slow readout.
Two different blocking filters were used to observe the targets (one on each M2FS channel), based on a visual examination of the VLT spectra.
For fainter targets with unclear Ba detections or upper limits, we used the \texttt{BulgeGC1} filter, which includes 24 fibers covering 6 orders from $6100-6700${\AA}, including the Ba line at 6496.7{\AA}. The Ba line at 6141{\AA} is on the blue end of the filter cutoff and cannot be used.
For brighter targets that already had clear Ba detections in the VLT data or upper limits, we instead used the \texttt{MgWide} filter, which includes 28 targets covering 4 orders from $5150-5400${\AA}.
6 sky fibers were allocated for each arm. The total exposure time was 14h.

The data were reduced with a custom pipeline\footnote{\url{https://github.com/alexji/m2fs_reduction}}.
Each of 4 amplifier images was bias subtracted using the overscan and stitched into one image, then had dark current subtracted.
Every science frame was associated with a single arc and flat obtained closest in time to the science frame.
The object trace was fit to each flat using a 5th order Legendre polynomial.
Scattered light was subtracted from every flat and science frame by fitting the inter-object regions with a 2D Legendre polynomial of degree 5 in either direction.
Twilight flats were used to determine throughput corrections for each fiber.

The wavelength calibration was motivated by \citet{Kelson03} and adapted for fiber spectroscopy.
An initial feature identification was done once by hand, extracting all orders of each fiber and using the IRAF \texttt{identify} command to manually identify positions of $50-70$ arc lines in each order of each fiber in the X (wavelength) direction on the CCD \citep{Tody1986,Tody1993}.
These identifications were then turned back into 2D coordinates using the trace functions.
The actual wavelength calibration was performed in 2D, finding sources in each arc frame using \texttt{Source\ Extractor} \citep{Bertin1996}, and matching the detected sources to identified lines using a KD tree\footnote{2D wavelength calibration was necessary for the \texttt{BulgeGC1} filter because the ThArNe arcs taken for this setting were extremely saturated, introducing many spurious features in 1D extracted arcs.}.
We then fit a 5th order Legendre polynomial for the wavelength solution, iteratively rejecting outliers, and using lines from all object fibers to fit the overall distortion.
A single X and Y pixel offset is allowed for each fiber (but not orders within a fiber) to account for any movement of the fibers in the pseudo-slit.
In total, $40-50$ lines were identified and used in each order for the \texttt{MgWide} filter, and $10-30$ lines were identified and used for the \texttt{BulgeGC1} filter (fewer due to the saturated arcs).
The final wavelength solution has a typical RMS ${<}0.01${\AA} in both arms.
Data were then extracted using flat-relative optimal extraction \citep{Zechmeister14}, which we found performed better than fitting a functional form to the object profile.

To perform sky subtraction, we linearly rebinned the extracted spectra onto a uniform wavelength grid, then followed essentially the same sky subtraction procedure as the VLT data.
The main differences were that the M2FS data has multiple orders, so sky subtraction was done independently for each order; and since the \texttt{MgWide} filter has few sky lines, we did not rescale the master sky spectrum to match line strengths, instead just directly subtracting the throughput-corrected master sky.
There were clear differences in the line spread function for different fibers, resulting in residuals around sky lines. We thus rejected data around sky lines.
Different exposures were coadded order-by-order with an inverse-variance weighted average.
Coadded orders were then continuum-normalized separately in \texttt{smhr}\footnote{\url{https://github.com/andycasey/smhr}, originally described in \citet{Casey14} and expanded in \citet{Ji20b}} before being stitched into a single spectrum.

\subsection{Magellan/M2FS MedRes}

We conducted two sets of medium resolution observations of Ret~II stars using M2FS
on 2016 September 6 and 10, totaling 6.72~hr of integration time.
We used the MedRes grating on the `R' spectrograph,
95~$\mu$m slits, 
2x2 binning, 
and 
the \texttt{MedRes\_Ba(23)} filter, which transmits one order
with $4450 \leq \lambda \leq 4615$~\AA.~
This setup yields $R \sim 9,000$, as measured
from individual Ar or Th emission lines in the comparison lamp spectra.
We performed data reduction, extraction, wavelength calibration,
sky subtraction, co-addition, and continuum normalization 
following the procedures described in \citet{Roederer16a},
modified for use with MedRes spectra.

We extracted the two sets of Ret~II MedRes spectra
(one set from each night)
separately.
We measured radial velocities of each star
in each observation by cross-correlating
(using the IRAF \texttt{fxcor} task)
its spectrum against a synthetic metal-poor template 
spectrum smoothed to the same spectral resolution.
Repeat observations of probable Ret~II members
show a standard deviation of 1.7~\kmsec, which we
regard as the uncertainty of an individual measurement.
We also observed two comparison stars, 
\object[CD -24 1782]{CD~$-$24$^{\circ}$1782},
and
\object[BPS CS 31082-001]{CS~31082--001} 
using the same M2FS MedRes setup.
Their radial velocities, 
after applying Heliocentric corrections
computed using the IRAF \texttt{rvcorrect} task,
agree with published values 
\citep{Roederer14c,Hansen15} 
to better than 1.7~\kmsec.
which we regard as the systematic uncertainty
of our measurements.
Combining the individual and systematic uncertainty, we adopt a total velocity uncertainty of 2.4~\kmsec\ for the MedRes velocities.
However, when later comparing repeat velocity measurements of Ret II stars, we found a systematic offset of ${\approx}$10~\kms. We thus decided not to use the M2FS MedRes velocities in this work, though we report their values in Appendix~\ref{app:rv}.

\subsection{Comment on Sky Subtraction and Ba Lines}\label{sec:skysub}

Our primary Ba line at 6496.7{\AA} is very close to a strong sky line at 6498.7{\AA}, and our Ba abundances are potentially susceptible to sky subtraction residuals.
For the VLT data with SNR $> 25$, we were able to obtain good simultaneous fits to the Ba lines and the sky line residuals. We decided stars with SNR $< 25$ were too strongly impacted by sky subtraction to have reliable Ba measurements, especially since a small error can make a big asymmetric difference in the abundances for saturated lines (Section~\ref{sec:vltfit}).
For the M2FS data, the signal-to-noise was lower and the heliocentric correction moved the Ba line right into the sky line, so none of the Ba 6496.7{\AA} line measurements from M2FS were reliable.
There was also a 6141{\AA} Ba line located on the filter cutoff, but after investigation we decided it was adversely affected by scattered light subtraction systematics and thus not sufficiently reliable for abundance measurements.

\begin{rotatetable*}
\begin{deluxetable*}{rllccrrrrrccrcrr}
\tablecolumns{16}
\setlength{\tabcolsep}{3pt}
\tabletypesize{\scriptsize}
\tablecaption{\label{tab:obs}Observations}
\tablehead{ID & RA & Dec & $g_0$ & $r_0$ & $r_e/r_{\rm h}$ & $v_{\rm hel}$ & $\sigma_v$ & $\mu_\alpha \cos \delta$ & $\mu_\delta$ & Mem & Bin & SNR/px & HiRes & SNR/px & SNR/px \\ & (h:m:s) & (d:m:s) & (mag) & (mag) & & (\kmsec) & (\kmsec) & (mas/yr) & (mas/yr) & & & (VLT) & Mode & (HiRes) & (MedRes)}
\startdata
  1 & 03:35:23.8 & -54:04:07.66 & 16.45 & 15.65 & 0.60 & $+65.3$ &  0.2 & $ 2.43$ & $-1.39$ & M & N &     140 & \nodata  & \nodata &      94 \\
  2 & 03:36:07.7 & -54:02:35.58 & 17.43 & 16.81 & 0.57 & $+62.5$ &  3.0 & $ 2.39$ & $-1.33$ & M & Y &      73 & MgWide   &      39 &      58 \\
  3 & 03:34:47.9 & -54:05:25.03 & 17.49 & 16.87 & 1.49 & $+62.0$ &  0.4 & $ 2.36$ & $-1.39$ & M & N &      94 & MgWide   &      34 &      57 \\
  4 & 03:35:31.1 & -54:01:48.24 & 17.61 & 17.02 & 0.79 & $+58.5$ &  0.3 & $ 2.25$ & $-1.35$ & M & N &      69 & MgWide   &      37 &      68 \\
  5 & 03:35:48.0 & -54:03:49.84 & 18.27 & 17.69 & 0.39 & $+61.7$ &  0.3 & $ 2.28$ & $-1.29$ & M & N &      76 & MgWide   &      19 &      40 \\
  6 & 03:35:37.1 & -54:04:01.25 & 18.57 & 18.03 & 0.37 & $+64.3$ &  0.4 & $ 2.50$ & $-1.60$ & M & N &      65 & MgWide   &      24 &      25 \\
  7 & 03:35:56.3 & -54:03:16.29 & 18.86 & 18.38 & 0.39 & $+63.2$ &  0.4 & $ 2.28$ & $-1.34$ & M & N &      50 & MgWide   &      14 &      26 \\
  8 & 03:34:57.6 & -54:05:31.42 & 18.94 & 18.40 & 1.25 & $+60.2$ &  0.4 & $ 2.34$ & $-1.28$ & M & N &      52 & MgWide   &      15 &      29 \\
  9 & 03:34:54.2 & -54:05:58.05 & 18.93 & 18.42 & 1.35 & $+69.2$ &  0.4 & $ 2.40$ & $-1.36$ & M & N &      52 & MgWide   &      17 &      33 \\
 10 & 03:35:21.0 & -54:03:48.16 & 18.93 & 18.43 & 0.68 & $+64.5$ &  4.6 & $ 2.24$ & $-1.20$ & M & Y &      53 & MgWide   &      15 & \nodata \\
 11 & 03:35:02.5 & -54:03:54.27 & 19.23 & 18.73 & 1.20 & $+67.0$ &  0.9 & $ 2.68$ & $-1.40$ & M & N & \nodata & BulgeGC1 &      12 &      30 \\
 12 & 03:35:58.1 & -54:02:04.78 & 19.30 & 18.81 & 0.27 & $+64.6$ &  0.5 & $ 2.35$ & $-1.13$ & M & N &      39 & MgWide   &      15 &      29 \\
 13 & 03:35:11.7 & -54:03:21.81 & 19.31 & 18.83 & 1.00 & $+67.0$ &  3.4 & $ 2.71$ & $-1.31$ & M & Y &      42 & MgWide   &      10 & \nodata \\
 14 & 03:34:39.7 & -54:07:54.37 & 19.37 & 18.84 & 1.82 & $+62.3$ &  0.7 & $ 2.54$ & $-1.51$ & C & N &      40 & BulgeGC1 &      10 & \nodata \\
 15 & 03:36:01.8 & -54:04:05.49 & 19.57 & 19.07 & 0.81 & $+62.6$ &  0.6 & $ 2.18$ & $-1.21$ & M & N &      34 & BulgeGC1 &      12 & \nodata \\
 16 & 03:35:50.1 & -54:01:39.24 & 19.72 & 19.29 & 0.39 & $+64.6$ &  0.9 & $ 2.47$ & $-1.63$ & C & N &      30 & MgWide   &      10 & \nodata \\
 17 & 03:35:13.7 & -54:04:56.72 & 19.67 & 19.20 & 0.87 & $+60.2$ &  0.7 & $ 2.52$ & $-1.30$ & M & N &      32 & BulgeGC1 &      12 &      27 \\
 18 & 03:35:17.0 & -54:04:03.05 & 19.68 & 19.20 & 0.77 & $+65.5$ &  8.0 & $ 2.52$ & $-1.08$ & M & Y &      21 & BulgeGC1 &      12 & \nodata \\
 19 & 03:35:15.2 & -54:08:43.03 & 19.69 & 19.21 & 1.81 & $+67.8$ & 11.1 & $ 2.39$ & $-1.07$ & M & Y &      31 & BulgeGC1 &      12 &      22 \\
 20 & 03:35:14.0 & -54:05:58.19 & 19.98 & 19.53 & 1.01 & $+63.6$ &  0.7 & $ 2.76$ & $-1.60$ & M & N &      27 & BulgeGC1 &       5 & \nodata \\
 21 & 03:36:35.8 & -54:01:20.21 & 20.19 & 19.73 & 1.23 & $+61.0$ &  0.8 & $ 2.32$ & $-1.37$ & M & N &      16 & BulgeGC1 &       7 & \nodata \\
 22 & 03:36:21.9 & -54:00:40.68 & 20.29 & 19.85 & 0.86 & $+65.6$ &  0.7 & $ 2.02$ & $-2.11$ & M & N &      15 & BulgeGC1 &       5 & \nodata \\
 23 & 03:35:24.0 & -54:02:26.69 & 20.28 & 19.85 & 0.82 & $+61.6$ &  0.7 & $ 2.70$ & $-1.78$ & M & N &      19 & BulgeGC1 &       6 & \nodata \\
 24 & 03:35:02.9 & -54:01:09.84 & 20.38 & 19.94 & 1.81 & $+63.6$ &  0.8 & $ 3.11$ & $-1.14$ & M & N &      19 & BulgeGC1 &       4 & \nodata \\
 25 & 03:35:44.2 & -54:01:50.03 & 20.41 & 19.93 & 0.43 & $+62.0$ &  0.7 & $ 2.22$ & $-2.30$ & M & N &      20 & BulgeGC1 &       6 & \nodata \\
 26 & 03:35:35.4 & -54:02:54.88 & 20.72 & 20.38 & 0.36 & $+61.5$ &  0.8 & $ 2.83$ & $-1.63$ & M & N &      13 & BulgeGC1 &       4 & \nodata \\
 97 & 03:36:27.7 & -53:58:26.31 & 17.13 & 16.27 & 1.34 & $+67.5$ &  0.7 & $ 2.34$ & $-1.29$ & C & N &     125 & \nodata  & \nodata & \nodata \\
 99 & 03:35:14.5 & -54:02:33.16 & 20.39 & 19.92 & 1.08 & $+67.2$ &  0.8 & $ 2.50$ & $-0.90$ & M & N &      19 & BulgeGC1 &       7 & \nodata \\
100 & 03:36:18.7 & -53:57:45.14 & 18.04 & 18.20 & 1.53 & $+61.0$ &  0.8 & $ 2.46$ & $-1.48$ & M & N & \nodata & MgWide   &      25 & \nodata \\
102 & 03:36:12.7 & -53:56:02.25 & 18.05 & 18.23 & 2.16 & $+66.8$ &  0.7 & $ 2.19$ & $-1.21$ & M & N & \nodata & MgWide   &      28 & \nodata \\
134 & 03:35:12.5 & -54:00:59.16 & 20.82 & 20.47 & 1.58 & $+62.2$ &  1.0 & $ 3.01$ & $-1.90$ & M & N &      13 & MgWide   &       2 & \nodata \\
142 & 03:35:38.7 & -54:04:55.59 & 20.89 & 20.42 & 0.67 & $+61.0$ &  0.7 & $ 2.36$ & $-1.46$ & C & N &      14 & MgWide   &       3 & \nodata \\
143 & 03:35:39.5 & -54:00:23.48 & 20.93 & 20.62 & 1.07 & $+58.9$ &  0.9 & $ 3.59$ & $-2.82$ & C & N &      10 & MgWide   &       3 & \nodata \\
144 & 03:35:39.4 & -54:03:57.31 & 20.93 & 20.67 & 0.35 & $+65.0$ &  1.1 & $ 2.06$ & $-1.47$ & M & N &      11 & MgWide   &       2 & \nodata \\
151 & 03:35:47.5 & -54:03:25.05 & 20.59 & 20.36 & 0.23 & $+68.0$ &  1.0 & $ 2.68$ & $-0.51$ & C & N &      14 & \nodata  & \nodata & \nodata \\
154 & 03:36:05.4 & -54:02:06.36 & 20.41 & 20.21 & 0.44 & $+61.9$ &  1.0 & $ 2.70$ & $-1.46$ & C & N &      13 & \nodata  & \nodata & \nodata \\
157 & 03:36:11.6 & -54:00:34.45 & 20.64 & 20.23 & 0.71 & $+67.9$ &  0.9 & $ 1.94$ & $-2.10$ & M & N &      12 & BulgeGC1 &       5 & \nodata \\
188 & 03:34:59.3 & -53:58:23.98 & 20.90 & 20.59 & 2.79 & $+68.4$ &  1.2 & $ 2.05$ & $-2.94$ & C & N &       7 & \nodata  & \nodata & \nodata \\
192 & 03:35:16.9 & -54:05:22.59 & 20.69 & 20.29 & 0.87 & $+69.1$ &  0.9 & $ 1.89$ & $-1.77$ & M & N &      17 & BulgeGC1 &       4 & \nodata \\
195 & 03:35:16.5 & -54:04:36.12 & 20.60 & 20.21 & 0.78 & $+68.1$ &  0.8 & $ 2.14$ & $-2.22$ & M & N &      14 & BulgeGC1 &       4 & \nodata \\
\enddata
\tablecomments{Positions and photometry from DES. Half-light radius assuming \citet{MutluPakdil18} structural parameters. Heliocentric radial velocities and uncertainties are the average systematics-corrected uncertainties described in Section~\ref{sec:rvmeasurement} and Appendix~\ref{app:rv}. Proper motions are from Gaia EDR3. The HiRes and MedRes columns refer to M2FS observations described in this paper. Only the member and candidate member stars are shown here. The full table is available in electronic form in the online journal.}
\end{deluxetable*}
\end{rotatetable*}

\section{Analysis}\label{sec:analysis}

We assigned each star a numerical ID. IDs $1-26$ were red giant branch stars identified as members or candidate members by \citet{Simon15} (a superset of \citealt{Walker15,Koposov15b}), sorted by magnitude.
IDs $97-200$ were assigned arbitrarily to other observed stars.

\subsection{Photometry and Astrometry}
For all observed stars, we queried the DES DR1 catalog for $griz$ magnitudes using NOAO datalab, which were dereddened using the DES DR1 reddening coefficients \citep{desdr1}.
The exception is star ID 1, which is saturated in the DES data release but had photometry determined from an individual DECam exposure in \citet{Simon15}.
The subscript $0$ (e.g., $g_0$) indicates dereddened magnitudes. 
RA and Dec were taken from the DES catalog (differing by ${\sim}$0\farcs1 compared to \Gaia).
We also computed the elliptical distance $r_e$ for each star, assuming Ret~II is centered at 03:35:47.83 $-$54:02:47.8 with a position angle of 68$^\circ$, ellipticity of 0.6, and half-light radius 6.3 arcmin \citep{MutluPakdil18}.
We adopted a distance modulus of $17.5$ whenever needed \citep{MutluPakdil18}.
Proper motions $\mu_\alpha \cos \delta$ and $\mu_\delta$ were obtained by cross-matching with \Gaia EDR3 \citep{Lindegren2021}.
Our spectroscopic target selection did not use proper motions, as it was performed before \Gaia DR2 was released.

\subsection{Radial Velocities} \label{sec:rvmeasurement}
Radial velocities were measured with weighted cross-correlation against a high-resolution template spectrum of \mbox{HD~122563} obtained with the MIKE spectrograph and shifted to rest-frame \citep{Bernstein03}.
Each spectrum was first normalized with a 3rd order polynomial.
Then, for a range of velocities, we shifted the template spectrum by that velocity and calculated the $\chi^2$ of pixels within a specific wavelength interval.
For the VLT data and the M2FS \texttt{BulgeGC1} data, we used H$\alpha$ to measure velocities, cross-correlating from $6550-6575$\AA.
For the M2FS \texttt{MgWide} data, we used the Mg triplet, cross-correlating from $5150-5200$\AA.
Velocities were then found as the minimum of the $\chi^2$ contour, and 1 $\sigma$ statistical velocity errors were determined by $\Delta\chi^2 = 1$  \citep[e.g.,][]{Martini2006}.
For the M2FS data, we used just one of the orders for velocity measurement: order 54 for \texttt{BulgeGC1} and order 69 for \texttt{MgWide}.
We did not attempt to determine very detailed radial velocities for the MedRes M2FS data, given its lower resolution.

We determined systematic velocity uncertainties from repeated velocity measurements.
For both the VLT and M2FS run, every star was observed with 7-15 individual exposures over two nights. We thus measured the velocity and statistical uncertainty for each individual exposure.
Then, following \citet{Li19S5}, we took all pairs of repeated velocity measurements with $\sigma_v < 30$ \kms and fit the following Gaussian-plus-outlier model:
\begin{equation}
\begin{split}
    v_i - v_j \sim &f \mathcal{N}\left(0, \sqrt{F(\sigma_{v,i})^2 + F(\sigma_{v,j})^2}\right) \\ + &(1-f) \mathcal{N}(0, \sigma_{\text{outlier}})
\end{split}
\end{equation}
where $v_i$, $v_j$, $\sigma_{v,i}$, and $\sigma_{v,j}$ are the individual velocities and uncertainties; $f$ is the fraction of pairs that are good; $\sigma_{\text{outlier}}$ is a large value characterizing a broad background outlier model; and $F(\sigma_v) = \sqrt{\sigma_{v,\text{floor}}^2 + (k \times \sigma_v)^2}$ is a rescaling of the velocity uncertainty with a scale factor $k$ and a systematic floor $\sigma_{v,\text{floor}}$.
The VLT data had $f=0.98$, $\sigma_{v, \text{floor}} = 0.69$ \kms, and $k=1.02$.
The M2FS \texttt{MgWide} data had $f=0.66$, $\sigma_{v, \text{floor}} = 0.74$ \kms, and $k=1.64$.
The M2FS \texttt{BulgeGC1} data had $f=0.93$, $\sigma_{v, \text{floor}} = 3.20\kms$, and $k=0.11$. In this case, having $k < 1$ suggests that the individual velocities are dominated by systematic errors, which is primarily due to the low SNR of individual exposures causing poor continuum fits and template matches. We thus conservatively take $k=1$ for the \texttt{BulgeGC1} data.
These values of $k$ and $\sigma_{v, \text{floor}}$ are applied to generate our final velocity uncertainties.

\begin{figure*}
    \centering
    \includegraphics[width=\linewidth]{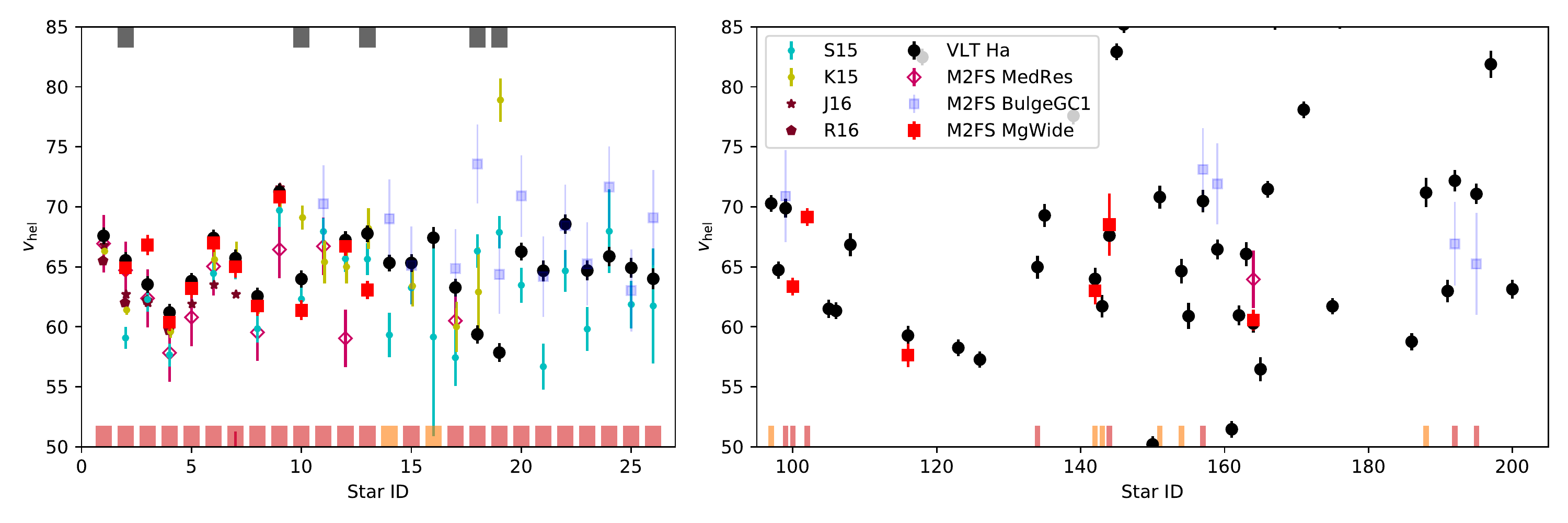}
    \caption{Individual radial velocity measurements for all stars from this study and the literature. No systematic velocity corrections have been applied in this figure.
    The left panel shows red giant branch stars \emph{previously} marked as confirmed or candidate members, while the right panel shows other stars. Along the bottom axis, red bars indicate stars that are certain members, while orange lines indicate candidate members. Along the top axis, grey lines indicate binary stars.}
    \label{fig:binaryvel}
\end{figure*}

Figure~\ref{fig:binaryvel} shows the individual velocity measurements for all observed stars.
The left panel shows the 26 likely members from previous observations, while the right panel shows other observed stars.
The red bars on the lower axis indicate stars that are clear members, while orange bars indicate candidate members (see Section~\ref{sec:membership} for the definitions).
For comparison, in the left panel we also plot velocities derived by \citetalias{Simon15}, \citetalias{Koposov15b}, \citetalias{Ji16c}, and \citetalias{Roederer16b}. Since many literature velocities were measured using the same spectra, duplicate velocity measurements from \citetalias{Walker15} and \citetalias{Simon15} were removed from this figure.

For the 40 members and candidate members (determined in Section~\ref{sec:membership}), we determined velocities by averaging our VLT and M2FS velocities with the literature velocities from \citetalias{Simon15}, \citetalias{Koposov15b}, \citetalias{Ji16c}, and \citetalias{Roederer16b}, for a total of up to 6 independent velocity measurements.
We combine all available literature velocities with an inverse-variance weighted mean. Details are given in Appendix~\ref{app:rv}, but this includes estimating a systematic velocity offset for each data sample using the \citetalias{Simon15} velocities as a reference, and identifying binary star candidates using a chi-squared test. Five candidate binary stars are identified in Table~\ref{tab:obs} and as grey bars on the top axis of Fig~\ref{fig:binaryvel}.
These binary star velocities were also found with a weighted mean, but the uncertainty indicates the range of velocities. Binary candidates are excluded from the velocity and velocity dispersion calculations.
Our focus in this paper is on chemical abundances, so this assessment of binarity and velocity systematics is incomplete, and a more comprehensive future study is warranted.

\subsection{Membership}\label{sec:membership}
We used four criteria to establish membership within our selected targets:
radial velocities in the range $53\,\kms < v_{\rm r, hel} < 74\,\kms$; 
proper motion within 2 units of Mahalanobis distance from the Ret~II mean proper motion of $(\mu_\alpha^*, \mu_\delta)=(2.39, -1.36)$ \citep{McConnachie2020}\footnote{For two vectors $x$ and $y$ with covariance matrix $\Sigma$, $d_{\text{Mahalanobis}} = \sqrt{(x-y)^\text{T} \Sigma^{-1} (x-y)}$.};
position within 2 elliptical half light radii of the Ret~II center\footnote{We used the structural parameters from \citet{MutluPakdil18}, but the membership is the same using the structural parameters from \citet{Munoz18}.};
and $g_0-r_0$ color between 13 Gyr, $\alpha$-enhanced Dartmouth isochrones \citep{Dotter08} with $\mbox{[Fe/H]}=-2.5$ and $\mbox{[Fe/H]}=-1.5$, allowing a 0.03 mag buffer on each side of the isochrone given an expected reddening uncertainty.
Since we used hard cuts, edge cases near these boundaries were inspected individually.

\begin{figure*}
    \centering
    \includegraphics[width=\linewidth]{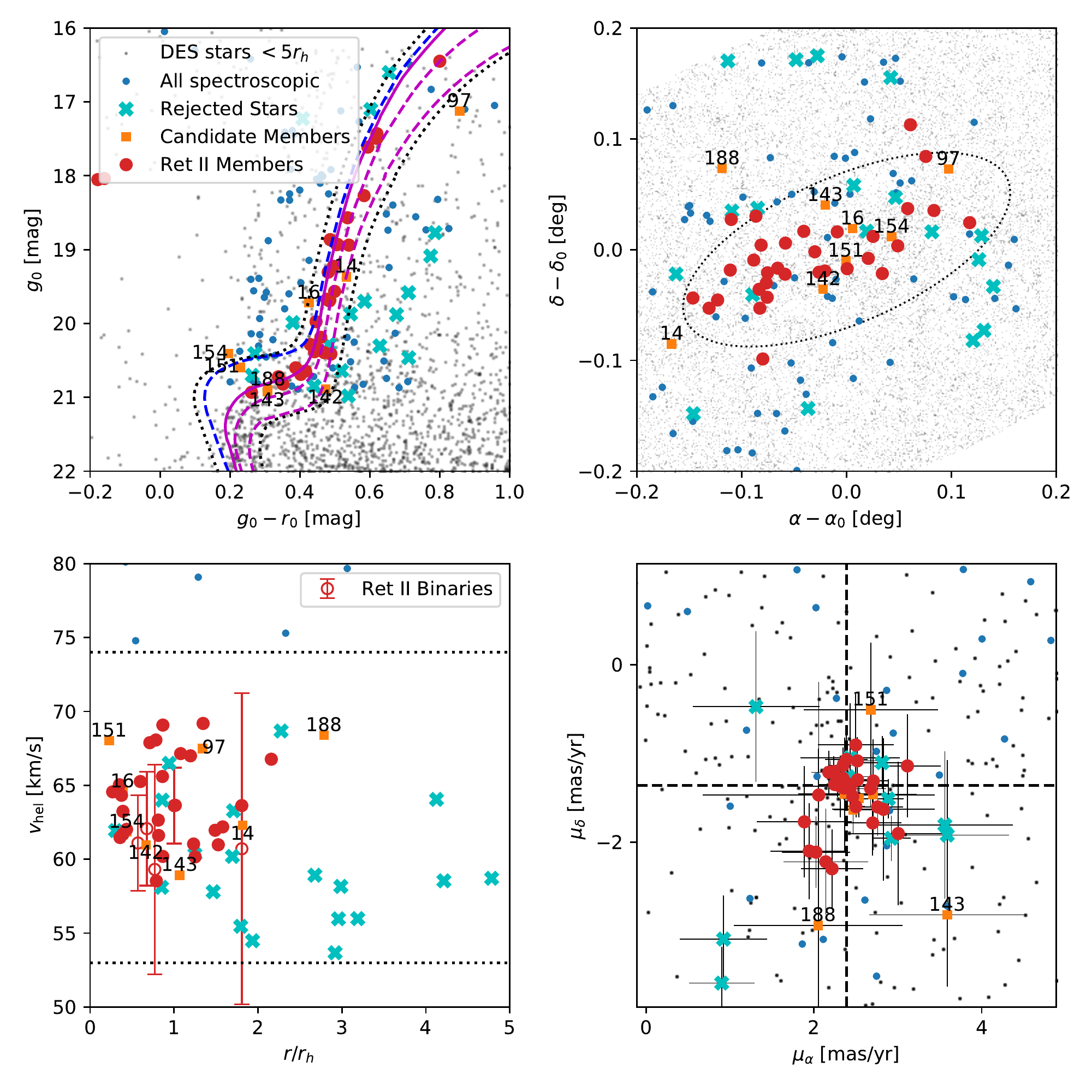}
    \caption{Diagnostic plots for membership determination. Red circles indicate clear Ret II members, orange squares are candidate Ret II members (with star ID labeled), cyan crosses are non-member stars with radial velocities similar to Ret II, small blue dots indicate spectroscopic targets with velocities far from Ret II, and small grey points indicate DES stars within 5 elliptical half-light radii. \emph{Top left:} color-magnitude diagram with dereddened DES photometry. The solid magenta line is a Dartmouth 13 Gyr, [Fe/H] $=-2.5$ isochrone. The dashed magenta lines are more metal-rich 13 Gyr isochrones with metallicities [Fe/H] $=-2.0$ and $-1.5$. The dashed blue line is a 10 Gyr isochrone with [Fe/H] $= -2.5$ to show where potential younger member stars or blue stragglers could lie.
    The dotted black lines show the region for membership, though individual stars outside this range were also considered.
    \emph{Top right:} spatial position of stars around the Ret~II center, only including stars within 5 elliptical half-light radii. The dotted black line shows 2 half light radii, which is used as the main membership cut although stars at larger radii were also considered.
    \emph{Bottom left:} radial velocity of spectroscopically observed stars vs elliptical half-light radius. Binaries are indicated as open circles, with the error bar spanning the range of observed velocities. There is no significant radial velocity gradient.
    The dotted black lines show the range of velocities allowed for membership.
    \emph{Bottom right:} \Gaia EDR3 proper motions. All our spectroscopic targets are bright enough to have EDR3 proper motions. Stars 14, 16, 97, 142, and 154 are located at the center and thus not labeled.
    Note the bulk of field stars are beyond the plot limits.
    The dashed black lines indicate the mean Ret~II proper motion.
    }
    \label{fig:membership}
\end{figure*}

The results are shown in Figure~\ref{fig:membership}.
70 of the 129 total spectroscopic targets were rejected as being outside of our radial velocity limits (small blue points).
Of the remaining 59 stars, 18 stars had consistent radial velocities but were rejected based on clearly discrepant spatial location, CMD position, or proper motion (cyan crosses). One additional star (ID 191) was also rejected as having [Fe/H] $\sim -0.5$ from our later analysis, too high to be part of Ret~II.
This leaves 40 stars, of which 32 stars were confident members, clearly matching at least three of the four criteria (large red points).
In the rest of this paper, we refer to these stars as ``clear members.''
The remaining 8 stars were highlighted as uncertain members (small orange squares), which we refer to as ``candidate members'' in the rest of this paper.
Stars 188 and 143 are CMD members but at somewhat large distance and proper motion away from the galaxy core to be considered very confident members.
Stars 14, 97, and 142 are redward of the isochrones, requiring high metallicities $\mbox{[Fe/H]}\gtrsim-1.5$ (or perhaps large carbon bands) to be considered part of Ret~II. Our subsequent analysis finds none of these stars has such a high metallicity.
Star 16 is blueward of the isochrone and has an unusually shallow H$\alpha$ line shape suggesting it is a hotter star.
Finally, stars 151 and 154 have consistent kinematics and would be CMD non-members, but they are located where evolved blue stragglers or an unusually young stellar population might be found, as indicated by a 10Gyr isochrone in the top-left panel of Figure~\ref{fig:membership}.
In other dwarf spheroidal galaxies, blue stragglers tend to appear even younger ($2.5 \pm 0.5$\,Gyr, \citealt{Santana2013}) so it is possible these stars could indicate a younger population of stars in Ret~II. However, the spectra of these potential member stars are too low SNR to reliable measure any abundances.
Stars 100 and 102 are known BHB member stars \citep{Simon15}.
Other than rejecting the clear non-member star ID 191, we have avoided using metallicity information in the membership selection so as to remain unbiased in final abundance distributions.

Note that several Ret~II members have slightly lower $\mu_\alpha$ and $\mu_\delta$ than most of the galaxy (at $(\mu_\alpha, \mu_\delta) = (+2.0, -2.0)$ compared to $(+2.4, -1.3)$). These 5 stars were also offset in the same direction in \Gaia\ DR2. However, they have larger proper motion uncertainties individually consistent with the bulk of the galaxy, and they do not stand out in radial velocity. Future \Gaia\ data releases can test whether these represent a true feature.

\subsection{Chemical Abundances}
We derived most abundances using a standard analysis using 1D \citet{Castelli04} model atmospheres and Local Thermodynamic Equilibrium (LTE) radiative transfer with MOOG \citep{Sneden73} including scattering \citep{Sobeck11} and \cite{Barklem00} damping\footnote{\url{https://github.com/alexji/moog17scat}}.
Since Ba is the element of most interest, we analyzed Ba with both LTE and non-LTE, see Section~\ref{sec:banlte}.
Our stellar parameters are given in Table~\ref{tab:sp},
atomic data in Table~\ref{tab:atomicdata},
and abundance results in Table~\ref{tab:abunds}.

\begin{deluxetable}{rrrrrrr}
\tablecolumns{7}
\tabletypesize{\footnotesize}
\tablecaption{\label{tab:sp}Stellar Parameters}
\tablehead{ID & $T_{\rm eff}$ & $\sigma_T$ & $\log g$ & $\sigma_g$ & $\nu_t$ & $\sigma_\nu$}
\startdata
  1 & 4655 &  53 & 1.24 & 0.19 & 1.61 & 0.14 \\
  2 & 4952 &  49 & 1.85 & 0.18 & 1.48 & 0.13 \\
  3 & 4953 &  49 & 1.87 & 0.18 & 1.48 & 0.13 \\
  4 & 5009 &  49 & 1.95 & 0.18 & 1.46 & 0.13 \\
  5 & 5034 &  50 & 2.23 & 0.18 & 1.41 & 0.13 \\
  6 & 5157 &  68 & 2.41 & 0.19 & 1.39 & 0.13 \\
  7 & 5316 &  77 & 2.61 & 0.19 & 1.36 & 0.13 \\
  8 & 5146 &  67 & 2.56 & 0.19 & 1.36 & 0.13 \\
  9 & 5248 &  68 & 2.60 & 0.19 & 1.36 & 0.13 \\
 10 & 5256 &  68 & 2.61 & 0.19 & 1.36 & 0.13 \\
 11 & 5286 &  71 & 2.74 & 0.19 & 1.34 & 0.13 \\
 12 & 5331 &  82 & 2.79 & 0.19 & 1.34 & 0.13 \\
 13 & 5336 &  83 & 2.80 & 0.19 & 1.33 & 0.13 \\
 14 & 5166 &  69 & 2.74 & 0.19 & 1.34 & 0.13 \\
 15 & 5275 &  70 & 2.87 & 0.19 & 1.33 & 0.13 \\
 16 & 5564 &  83 & 3.06 & 0.19 & 1.31 & 0.13 \\
 17 & 5349 &  88 & 2.95 & 0.19 & 1.32 & 0.13 \\
 18 & 5320 &  77 & 2.94 & 0.19 & 1.32 & 0.13 \\
 19 & 5324 &  79 & 2.94 & 0.19 & 1.32 & 0.13 \\
 20 & 5481 &  92 & 3.13 & 0.20 & 1.30 & 0.13 \\
 21 & 5420 &  93 & 3.18 & 0.20 & 1.30 & 0.13 \\
 22 & 5492 &  91 & 3.26 & 0.19 & 1.29 & 0.13 \\
 23 & 5550 &  86 & 3.28 & 0.19 & 1.29 & 0.13 \\
 24 & 5499 &  91 & 3.30 & 0.19 & 1.29 & 0.13 \\
 25 & 5309 &  75 & 3.22 & 0.19 & 1.29 & 0.13 \\
 26 & 5915 &  88 & 3.60 & 0.19 & 1.27 & 0.13 \\
 97 & 4584 &  64 & 1.47 & 0.19 & 1.56 & 0.14 \\
 99 & 5379 &  95 & 3.25 & 0.20 & 1.29 & 0.13 \\
134 & 5861 &  91 & 3.62 & 0.19 & 1.27 & 0.13 \\
142 & 5363 &  90 & 3.44 & 0.20 & 1.28 & 0.13 \\
143 & 6055 &  91 & 3.73 & 0.19 & 1.26 & 0.13 \\
144 & 6263 & 113 & 3.81 & 0.20 & 1.26 & 0.13 \\
151 & 6439 & 124 & 3.74 & 0.20 & 1.26 & 0.13 \\
154 & 6645 & 130 & 3.73 & 0.20 & 1.26 & 0.13 \\
157 & 5602 &  79 & 3.44 & 0.19 & 1.28 & 0.13 \\
188 & 6046 &  92 & 3.72 & 0.19 & 1.26 & 0.13 \\
192 & 5650 &  76 & 3.48 & 0.19 & 1.27 & 0.13 \\
195 & 5704 &  87 & 3.47 & 0.19 & 1.27 & 0.13 \\
\enddata
\tablecomments{We adopt $\mbox{[M/H]} = -2.5$ and $\mbox{[$\alpha$/Fe]}=+0.4$ for all stars.}
\end{deluxetable}

\begin{deluxetable}{rrrrl}
\tablecolumns{5}
\tablecaption{\label{tab:atomicdata}Atomic Data}
\tablehead{$\lambda$ (\AA) & Species & $\chi$ (eV) & $\log gf$ & Reference}
\startdata
\cutinhead{VLT}
   6439.075 &     Ca I & 2.524 &  0.390 & SR81          \\
   6493.781 &     Ca I & 2.521 & -0.109 & SR81          \\
   6494.980 &     Fe I & 2.402 & -1.273 & BLA82         \\
   6496.897 &    Ba II & 0.604 & -0.380 & GAL20 \\
\cutinhead{M2FS Mg Wide}
   5171.596 &     Fe I & 1.484 & -1.720 & OBR91 \\
   5172.684 &     Mg I & 2.712 & -0.393 & NIST                \\
   5183.604 &     Mg I & 2.717 & -0.167 & NIST                \\
   5269.537 &     Fe I & 0.858 & -1.330 & OBR91  \\
   5328.532 &     Fe I & 1.556 & -1.850 & OBR91  \\
   5371.489 &     Fe I & 0.957 & -1.640 & OBR91  \\
\cutinhead{M2FS MedRes}
   4554.03  &     56.1 & 0.000 & +0.17  & IVA06 \\
\enddata
\tablecomments{References: SR81 \citep{Smith1981}, BLA82 \citep{Blackwell1982}, IVA06 \citep{Ivans06}, GAL20 \citep{Gallagher2020}, OBR91 \citep{Obrian91Fe}, NIST \citep{NIST}, accessed through the Kurucz, VALD3, and linemake databases \citep{Kurucz95,VALD,Placco2021}.}
\end{deluxetable}

\begin{deluxetable*}{r|rrr|rr|rr|rr|rrr|rr}
\setlength{\tabcolsep}{3pt}
\tablecolumns{15}
\tablecaption{\label{tab:abunds}Abundances}
\tablehead{ID & Mem & VLT SNR & M2FS SNR & [Mg/H] & $\sigma_{\rm Mg}$ & [Ca/H] & $\sigma_{\rm Ca}$ & [Fe/H] & $\sigma_{\rm Fe}$ & [Ba/H]$_{\rm NLTE}$ & $\sigma_{\rm Ba}$ & [Ba/H]$_{\rm LTE}$ & [Ba/H]$_{\rm MR}$ & $\sigma_{\rm Ba,MR}$}
\startdata
  1 & M &     140 & \nodata & \nodata & \nodata & $-2.69$ & $ 0.07$ & $-2.89$ & $ 0.10$ & $-1.86$ & $ 0.18$ & $-1.58$ & $-1.64$ & $ 0.18$ \\
  2 & M &      73 & Mgb, 39 & $-2.48$ & $ 0.12$ & $-2.39$ & $ 0.06$ & $-2.76$ & $ 0.09$ & $-1.63$ & $ 0.17$ & $-1.33$ & $-1.13$ & $ 0.16$ \\
  3 & M &      94 & Mgb, 34 & $-2.59$ & $ 0.13$ & $-2.46$ & $ 0.06$ & $-2.77$ & $ 0.08$ & $-1.58$ & $ 0.17$ & $-1.28$ & $-1.42$ & $ 0.25$ \\
  4 & M &      69 & Mgb, 37 & $-2.55$ & $ 0.11$ & $-2.95$ & $ 0.07$ & $-2.66$ & $ 0.19$ & $-3.46$ &   limit & $-3.46$ & $-0.83$ &   limit \\
  5 & M &      76 & Mgb, 19 & $-2.52$ & $ 0.13$ & $-1.87$ & $ 0.07$ & $-2.02$ & $ 0.13$ & $-1.86$ & $ 0.15$ & $-1.57$ & $-1.60$ & $ 0.34$ \\
  6 & M &      65 & Mgb, 24 & $-2.75$ & $ 0.13$ & $-2.91$ & $ 0.08$ & $-3.18$ & $ 0.12$ & $-1.83$ & $ 0.18$ & $-1.54$ & $-1.04$ & $ 0.29$ \\
  7 & M &      50 & Mgb, 14 & $-2.67$ & $ 0.22$ & $-2.77$ &   limit & $-2.93$ &   limit & $-1.64$ &   limit & $-1.64$ & $-0.83$ &   limit \\
  8 & M &      52 & Mgb, 15 & $-2.03$ & $ 0.16$ & $-2.19$ & $ 0.20$ & $-2.14$ & $ 0.15$ & $-1.35$ & $ 0.17$ & $-1.12$ & $-0.98$ & $ 0.22$ \\
  9 & M &      52 & Mgb, 17 & $-2.72$ & $ 0.13$ & $-2.98$ & $ 0.08$ & $-2.89$ & $ 0.14$ & $-1.38$ & $ 0.35$ & $-1.08$ & $-1.14$ & $ 0.29$ \\
 10 & M &      53 & Mgb, 15 & $-2.09$ & $ 0.18$ & $-2.88$ & $ 0.08$ & $-3.28$ &   limit & $-3.41$ &   limit & $-3.41$ & \nodata & \nodata \\
 11 & M & \nodata & H$\alpha$, 12 & \nodata & \nodata & \nodata & \nodata & \nodata & \nodata & \nodata & \nodata & \nodata & $-1.14$ & $ 0.37$ \\
 12 & M &      39 & Mgb, 15 & $-2.84$ & $ 0.15$ & $-2.69$ & $ 0.11$ & $-2.72$ &   limit & $-1.50$ & $ 0.38$ & $-1.20$ & $-1.16$ & $ 0.37$ \\
 13 & M &      42 & Mgb, 10 & $-2.60$ & $ 0.16$ & $-2.11$ &   limit & $-2.88$ & $ 0.16$ & $-1.64$ & $ 0.22$ & $-1.35$ & \nodata & \nodata \\
 14 & C &      40 & H$\alpha$, 10 & \nodata & \nodata & $-2.84$ & $ 0.09$ & $-3.14$ & $ 0.20$ & $-2.59$ & $ 0.25$ & $-2.53$ & \nodata & \nodata \\
 15 & M &      34 & H$\alpha$, 12 & \nodata & \nodata & $-2.64$ &   limit & $-3.07$ &   limit & $-2.83$ &   limit & $-2.83$ & \nodata & \nodata \\
 16 & C &      30 & Mgb, 10 & \nodata & \nodata & $-1.94$ &   limit & $-2.99$ &   limit & $-0.19$ &   limit & $-0.19$ & \nodata & \nodata \\
 17 & M &      32 & H$\alpha$, 12 & \nodata & \nodata & $-2.33$ & $ 0.09$ & $-2.60$ & $ 0.21$ & $-1.84$ & $ 0.21$ & $-1.63$ & $-1.32$ & $ 0.47$ \\
 18 & M &      21 & H$\alpha$, 12 & \nodata & \nodata & $-1.93$ &   limit & $-2.52$ &   limit & $+0.10$ &   limit & $+0.10$ & \nodata & \nodata \\
 19 & M &      31 & H$\alpha$, 12 & \nodata & \nodata & $-2.72$ & $ 0.12$ & $-2.66$ & $ 0.21$ & $-1.61$ & $ 0.23$ & $-1.35$ & $-1.82$ & $ 0.87$ \\
 20 & M &      27 & H$\alpha$, 5 & \nodata & \nodata & $-1.66$ &   limit & $-2.21$ &   limit & $+0.69$ &   limit & $+0.69$ & \nodata & \nodata \\
 21 & M &      16 & H$\alpha$, 7 & \nodata & \nodata & $-1.84$ &   limit & $-1.74$ &   limit & $-0.49$ &   limit & $-0.49$ & \nodata & \nodata \\
 22 & M &      15 & H$\alpha$, 5 & \nodata & \nodata & $-1.32$ &   limit & $-2.22$ &   limit & $-0.19$ & $ 0.53$ & $-0.21$ & \nodata & \nodata \\
 23 & M &      19 & H$\alpha$, 6 & \nodata & \nodata & $-1.52$ &   limit & $-2.00$ &   limit & $+0.79$ &   limit & $+0.79$ & \nodata & \nodata \\
 24 & M &      19 & H$\alpha$, 4 & \nodata & \nodata & $-1.23$ &   limit & $-1.86$ &   limit & $+0.46$ &   limit & $+0.46$ & \nodata & \nodata \\
 25 & M &      20 & H$\alpha$, 6 & \nodata & \nodata & $-2.38$ & $ 0.20$ & $-2.13$ & $ 0.44$ & $-0.56$ & $ 0.56$ & $-0.60$ & \nodata & \nodata \\
 26 & M &      13 & H$\alpha$, 4 & \nodata & \nodata & \nodata &   limit & $-2.21$ &   limit & $-1.55$ & $ 0.27$ & $-1.35$ & \nodata & \nodata \\
 97 & C &     125 & \nodata & \nodata & \nodata & $-2.36$ & $ 0.11$ & $-2.35$ & $ 0.16$ & $-1.84$ & $ 0.17$ & $-1.64$ & \nodata & \nodata \\
 99 & M &      19 & H$\alpha$, 7 & \nodata & \nodata & \nodata &   limit & $-2.30$ & $ 0.27$ & $-0.63$ & $ 0.50$ & $-0.60$ & \nodata & \nodata \\
100 & M & \nodata & Mgb, 25 & \nodata & \nodata & \nodata & \nodata & \nodata & \nodata & \nodata & \nodata & \nodata & \nodata & \nodata \\
102 & M & \nodata & Mgb, 28 & \nodata & \nodata & \nodata & \nodata & \nodata & \nodata & \nodata & \nodata & \nodata & \nodata & \nodata \\
134 & M &      13 &  Mgb, 2 & \nodata & \nodata & $-1.62$ &   limit & $-1.90$ &   limit & $+0.68$ &   limit & $+0.68$ & \nodata & \nodata \\
142 & C &      14 &  Mgb, 3 & \nodata & \nodata & $-1.89$ & $ 0.23$ & $-1.98$ &   limit & $-0.64$ & $ 0.71$ & $-0.68$ & \nodata & \nodata \\
143 & C &      10 &  Mgb, 3 & \nodata & \nodata & $-0.86$ &   limit & $-1.92$ &   limit & $-1.32$ & $ 0.47$ & $-1.04$ & \nodata & \nodata \\
144 & M &      11 &  Mgb, 2 & \nodata & \nodata & $-1.65$ &   limit & $-1.27$ &   limit & $-0.14$ &   limit & $-0.14$ & \nodata & \nodata \\
151 & C &      14 & \nodata & \nodata & \nodata & $-1.58$ &   limit & $-1.64$ &   limit & $+0.36$ &   limit & $+0.36$ & \nodata & \nodata \\
154 & C &      13 & \nodata & \nodata & \nodata & $ 0.34$ &   limit & $-0.37$ &   limit & $-0.34$ & $ 0.61$ & $+0.23$ & \nodata & \nodata \\
157 & M &      12 & H$\alpha$, 5 & \nodata & \nodata & $-1.45$ &   limit & $-1.21$ &   limit & $+0.71$ &   limit & $+0.71$ & \nodata & \nodata \\
188 & C &       7 & \nodata & \nodata & \nodata & $-0.26$ &   limit & $-1.09$ &   limit & $+0.30$ &   limit & $+0.30$ & \nodata & \nodata \\
192 & M &      17 & H$\alpha$, 4 & \nodata & \nodata & $-1.65$ &   limit & $-2.11$ &   limit & $+0.12$ &   limit & $+0.12$ & \nodata & \nodata \\
195 & M &      14 & H$\alpha$, 4 & \nodata & \nodata & $-2.22$ &   limit & $-2.02$ &   limit & $-1.56$ & $ 0.37$ & $-1.34$ & \nodata & \nodata \\
\enddata
\tablecomments{In column ``Mem'', M indicates a clear member, and C indicates a candidate member. In the abundance uncertainty columns, ``limit'' means the indicated value is a 99\% upper limit. [Ba/H]$_{\rm NLTE}$ refers to our fiducial Ba abundances derived from MULTI/MARCS, [Ba/H]$_{\rm LTE}$ is the Ba abundances derived from MOOG/ATLAS, and [Ba/H]$_{\rm MR}$ is the MOOG/ATLAS abundances from the 4554{\AA} line derived from the medium-resolution M2FS observations.}
\end{deluxetable*}

\subsubsection{Stellar Parameters}
Since our spectra cover a very small wavelength range with few lines, we determined effective temperatures for member stars from DES photometry and Dartmouth isochrones \citep{Dotter08}.
We adopted a 13 Gyr, $\mbox{[Fe/H]}=-2.5$, [$\alpha$/Fe]$=+0.4$ isochrone as our fiducial isochrone. The isochrone was used to fit $\Teff$ as a function of $g_0-r_0$, and then we applied this to every member star.
The brightest known member star (ID=1) is near the DES saturation limit with obviously incorrect $r$ and $i$ magnitudes in DES DR1. We adopted $g_0-r_0=0.80$ from \citet{Simon15} to determine this star's stellar parameters.
Statistical uncertainties were calculated assuming a fixed error in $g_0-r_0 = 0.02$ mag (the typical reddening uncertainty) resulting in a typical temperature error of 50--100\,K.
Systematic uncertainties were estimated by taking the largest difference between the fiducial isochrone and the $\Teff$ calculated using isochrones of (12, 13, 14) Gyr, $\mbox{[Fe/H]} = (-2.5, -2.0)$, and $\mbox{[$\alpha$/Fe]} = (0.0, 0.4)$, a typical uncertainty of 30--40\,K.
The total temperature uncertainty was the quadrature sum of these two uncertainties.

The surface gravity $\logg$ was determined photometrically using the equation $\logg = 4.44 + \log M_\star + 4 \log \Teff/5780\text{K} + 0.4 (g_0 - \mu + BC(g) - 4.75)$ \citep{Venn17} where $M_\star = 0.75 \pm 0.1 M_\odot$ is the typical mass of an old red giant branch star, $g_0$ is the dereddened DES $g$ magnitude, $\mu = 17.5$ is the distance modulus to Ret II, and $BC(g)$ is the \citet{Casagrande14} bolometric correction.
Besides the $M_\star$ and temperature uncertainties, we propagated a conservative 0.2 mag total uncertainty for the distance modulus and dereddening, as well as 0.03 mag uncertainty in the bolometric correction.
The total $\logg$ uncertainty is about 0.2 dex for all stars.
Using $r$ instead of $g$ led to the same $\logg$ within 0.05 dex, and using different metallicities for the bolometric correction led to differences within 0.02 dex.

The $\Teff$ and $\logg$ in Table~\ref{tab:sp} agree with two previous works that studied a total of 9 stars in Ret~II using high-resolution spectroscopy \citepalias{Ji16c,Roederer16b}. The stellar parameters in these two studies were derived using standard 1D-LTE methods, i.e., by balancing abundances of Fe lines with respect to excitation potential, ionization, and line strength, including a temperature recalibration to a photometric scale \citep{Frebel13}. All nine stars agree within the $1\sigma$ stellar parameter uncertainties with no mean offset.

Our primary line of interest, the 6496{\AA} Ba line, is saturated in essentially all of our stars with a detected Ba line.
Microturbulence ($\nu_t$) thus plays a major role in determining the final abundance and uncertainty, especially for the two coolest stars in our sample (IDs 1 and 97).
As we do not have enough lines to determine $\nu_t$ self-consistently in our stars, we instead adopt a microturbulence relation based on $\log g$ from the metal-poor giants in \citet{Roederer14c}:
\begin{equation}
    \nu_t = 0.039 (\log g)^2 - 0.331 (\log g) + 1.960
\end{equation}
where the typical scatter around the relation is 0.13 km\,s$^{-1}$.
We note that this relation is quite different from the direct $\nu_t$ measurements in \citetalias{Ji16c} and \citetalias{Roederer16b}, but it eliminates a trend in Ba abundance with effective temperature that would otherwise be present.
We have decided to adopt this relation for the rest of the main paper, and a complete investigation of microturbulence choices and the impact on our results is discussed in Appendices~\ref{app:vtba}~and~\ref{app:basys}.

As we do not have Fe or $\alpha$-element constraints for most stars, we assume everywhere a model metallicity of $\mbox{[M/H]} = -2.5$ and $\mbox{[$\alpha$/Fe]}=+0.4$. We verified in stars with $\mbox{[Fe/H]}$ and $\mbox{[Mg/Fe]}$ measurements that changing these parameters produces negligible differences to the resulting abundances.
The final stellar parameters and uncertainties for all members or candidate members are given in Table~\ref{tab:sp}.
We do not determine stellar parameters for the BHB stars (IDs 100 and 102).

\subsubsection{VLT Data}\label{sec:vltfit}
In the VLT data, most Ret II member stars only have 1-5 significant absorption features:
H-$\alpha$, the Ba~\textsc{ii} line at 6496.897{\AA}, an Fe~\textsc{i} line at 6494.98{\AA}, a Ca~\textsc{i} line at 6439.08{\AA}, and sometimes a Ca~\textsc{i} line at 6493.78{\AA}.
We use this data to derive Ba, Fe, and Ca abundances where possible. Atomic data used are given in Table~\ref{tab:atomicdata}.

\begin{figure*}
    \centering
    \includegraphics[width=\linewidth]{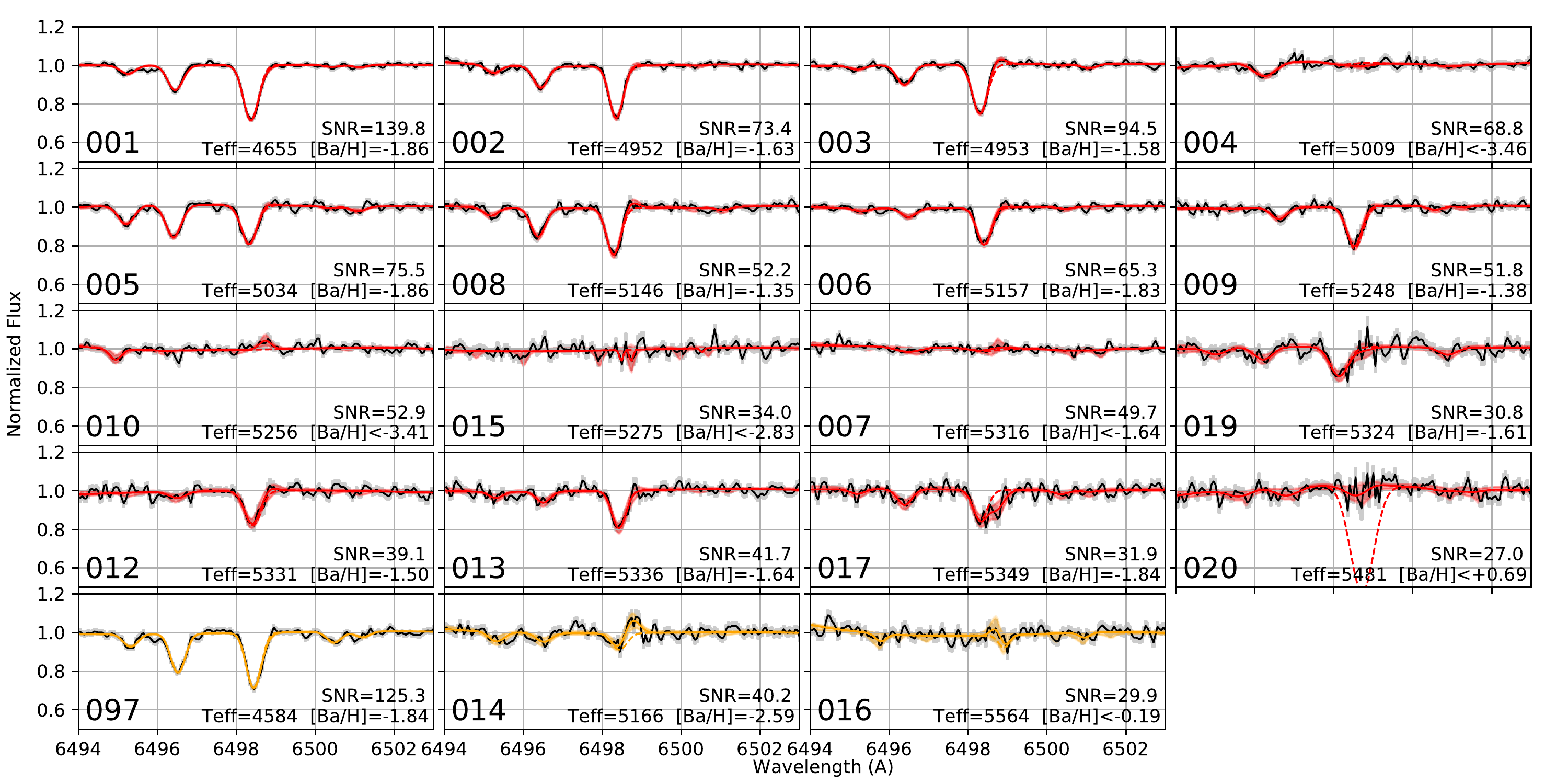}
    \includegraphics[width=\linewidth]{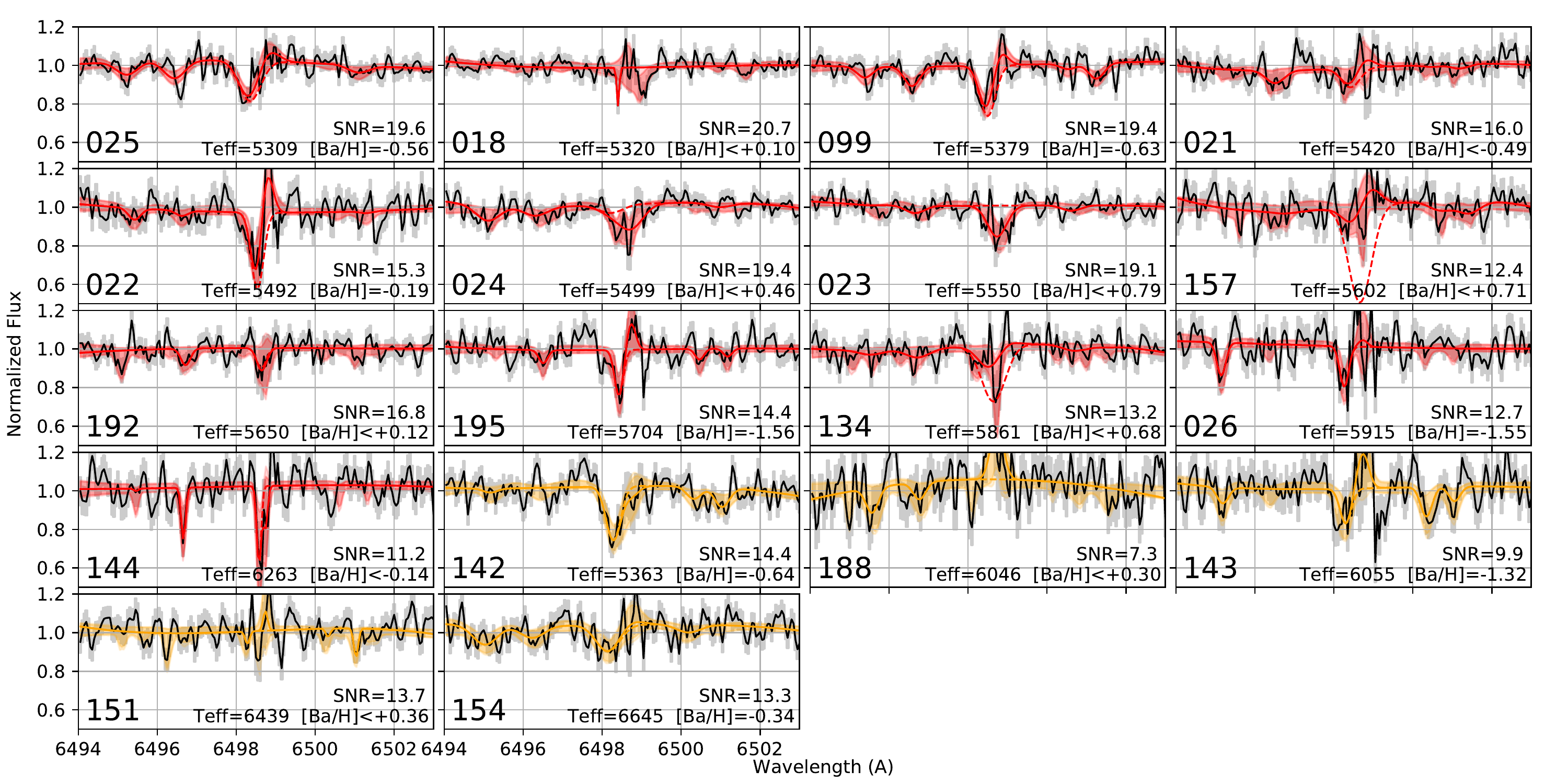}
    \caption{
    VLT spectra with best-fit models of the Ba line.
    From low to high wavelength the primary absorption features are due to Ca, Fe, and Ba.
    The Ba abundance measurement or limit is indicated.
    Stars are sorted by SNR, membership, and effective temperature.
    Top section: stars with SNR $> 25$. Bottom section: stars with SNR $< 25$.
    The sky residuals are clearly substantial for the lower SNR stars.
    The black line shows the data, while the grey shaded region indicates the 1$\sigma$ spectrum noise.
    Best-fit models are shown in red and orange for clear and candidate members, respectively.
    The solid colored lines are the model including the sky line, while the dashed colored lines are the model fit in the absence of the fitted sky line residual. In some cases (e.g. star 020, 134, 157) the Ba line is completely degenerate with the sky emission line, so no useful constraint is obtained.
    The shaded colored regions indicate the 16-84th percentile range of model parameters.
    Note that we have not explicitly plotted the upper limit model.
    }
    \label{fig:vltfit}
\end{figure*}

The Ba and Fe abundances are derived by fitting the spectral region from 6494$-$6504{\AA}.
Given the systemic velocity of 64 \kms, the Ba line is right next to a strong and variable sky line at 6498.74\,{\AA}.
The sky subtraction procedure described in Section~\ref{sec:obsred} often leaves a significant residual (Figure~\ref{fig:vltfit}). After testing several possible alternate sky subtraction procedures, we decided that the best course of action was to include the sky line residual in the model of this spectral region and marginalize over the uncertainty in our final results.
For all member stars, a coarse normalization was first performed using a sigma-clipped 3rd order polynomial between 6450$-$6550{\AA}. We then excised a region of the spectrum from 6494$-$6504{\AA} to fit in detail. Our model of this region is an 8$+$4 parameter model summarized in Table~\ref{tab:params}: a single Gaussian line width, six amplitudes for the Ca, Fe, and Ba stellar absorption features, a residual amplitude characterizing the sky line, a linear wavelength shift applied to the star but not the sky line, and a 3rd-order polynomial to fit the residual continuum.
Using the same line width for both stellar and sky features implicitly assumes that all lines were unresolved by the spectrograph, which is valid here.
Note that the two lines at 6498.9{\AA} and 6499.7{\AA} are not present in any of the Ret II stars, but these were included in the model to fit more metal-rich foreground stars.

The model parameters were optimized using \texttt{scipy.optimize.curve\_fit}.
For five stars with high SNR (IDs 1, 3, 4, 5, 97), the best-fit $\chi^2$ was larger than 1, suggesting that the data uncertainties were underestimated or the model did not provide a sufficient description of the data. For these stars, we increase the data uncertainties by 5--30\% such that the reduced $\chi^2$ is 1. The other stars had reduced $\chi^2$ of ${\approx} 0.8-0.9$, but we leave their errors unscaled.
Uncertainties in the fit were then found using dynamic nested sampling with \texttt{dynesty} \citep{dynesty}.
The priors used for the sampling are listed in Table~\ref{tab:params}.
Every fit and MCMC chain was visually inspected to ensure goodness of fit before accepting its results.
Lines considered poor visual fits were marked for upper limit determination.

LTE abundances and uncertainties were determined by putting equivalent width distributions into curves of growth calculated by MOOG.
First, the posterior distribution of the absorption line parameters were analytically converted into a posterior for the equivalent widths that marginalized over the effect of the sky line.
Curves of growth were constructed with MOOG, which were used to convert the equivalent width distribution to an abundance distribution.
For Ba, the effect of hyperfine structure was included assuming an $r$-process isotope distribution \citep{Ivans06,Sneden08}.
For detected lines, we adopt the optimum fit as the point estimate and larger of the difference between the point estimate and the 16th and 84th percentiles as a $1\sigma$ abundance error.
For undetected lines, we adopt the 99th percentile of the abundance posterior as the upper limit.
Despite this effort, stars with SNR $< 25$ still appear to have unreliable Ba abundances (Figure~\ref{fig:vltfit}, Appendix~\ref{app:vtba}).

\begin{deluxetable}{ccc}
    \tablecolumns{3}
    \tablecaption{\label{tab:params}VLT Spectrum Fit Parameters}
    \tablehead{Parameter & Description & Prior}
    \startdata
    $A_{\text{Ba}}$       & Amplitude of 6496.9\AA\ Ba line & $\mathcal{U}[0, 1]$ \\
    $\lambda_{\text{Ba}}$ & Observed wavelength of Ba line & $\mathcal{U}[6493.4, 6503.4]$ \\
    $\sigma$              & Width of all lines & $\mathcal{U}[0.01, 0.30]$ \\
    $A_{\text{sky}}$      & Amplitude of 6498.7\AA\ sky line & $\mathcal{U}[-0.50, 0.50]$ \\
    $A_{\text{Fe}}$       & Amplitude of 6495.0\AA\ Fe line & $\mathcal{U}[0, 1]$ \\
    $A_{\text{Ca}}$       & Amplitude of 6493.8\AA\ Ca line & $\mathcal{U}[0, 1]$ \\
    $A_{\text{Fe,2}}$     & Amplitude of 6498.9\AA\ Fe line & $\mathcal{U}[0, 1]$ \\
    $A_{\text{Ca,2}}$     & Amplitude of 6499.7\AA\ Ca line & $\mathcal{U}[0, 1]$ \\
    $c_0$     & Constant Continuum Coefficient & $\mathcal{U}[0.5, 1.5]$ \\
    $c_{1,2,3}$     & Continuum Coefficient & $\mathcal{U}[-1.0, 1.0]$ \\
    \enddata
\end{deluxetable}

Stellar parameter uncertainties were found by calculating new curves of growth for $1\sigma$ differences in stellar parameters, redoing the above calculations, and taking the difference.
These are added in quadrature to the statistical uncertainties.

As part of the above procedure we fit the Ca line at 6493.8{\AA}. However, there is a stronger Ca line at 6439{\AA} that is more often detected and also less susceptible to non-LTE effects \citep{Mashonkina17Ca}.
The Ca abundance is thus derived using this 6439{\AA} line with \texttt{smhr}. This includes the normalization, equivalent width measurement, and stellar parameter uncertainty propagation \citep{Ji20b}.
Formal 4 $\sigma$ upper limits for Ca were also calculated in \texttt{smhr} by synthesizing a Ca line such that $\Delta \chi^2 = 16$.

\subsubsection{M2FS HiRes Abundances}
We used \texttt{smhr} to normalize and stitch coadded orders, fit equivalent widths, measure element abundances from MOOG, and propagate stellar parameter uncertainties (see description in \citealt{Casey14,Ji20b}).
The primary use of this data is to measure the Mg abundances, which we derive from equivalent widths of the Mg b lines at 5172 and 5183{\AA}.
Useful abundances were ultimately only derived for the \texttt{MgWide} arm, since the \texttt{BulgeGC1} arm only had fainter and warmer stars, and the 6496{\AA} Ba line was completely blended with a sky line.

\subsubsection{M2FS MedRes Abundances}
The primary line of interest in these data is the Ba 4554{\AA} line.
We measure the abundance from this line with a procedure identical to that used for the VLT data, except that when fitting the equivalent width we manually define valid continuum wavelength ranges instead of modeling the many absorption lines. The results are provided in Table~\ref{tab:abunds}, and they are consistent with the Ba abundances derived from the 6496{\AA} line in the VLT data but with larger uncertainties. We thus do not use these abundances further.
However, in the future this medium-resolution mode may be an efficient way to search for strong Ba lines in other UFDs.

\subsubsection{Barium NLTE Abundances} \label{sec:banlte}

Ba can be substantially affected by non-LTE effects \citep[NLTE, e.g.,][]{Bergemann14,Mashonkina14,Gallagher2020} especially when the lines are near saturation. For metal-poor giants, NLTE makes the Ba 6497 line stronger, resulting in lower inferred Ba abundances when including NLTE.
To determine this quantitatively, we computed NLTE abundances for the Ba 6497 line using an updated version of the MULTI radiative transfer code \citep{MULTI,Bergemann19,Gallagher2020} and MARCS model atmospheres \citep{Gustafsson2008}.
Like the MOOG/ATLAS analysis, we adopted a metallicity of [M/H] $=-2.5$ and [$\alpha$/Fe] $=+0.4$ for all model atmospheres, and used the model atom presented in \citet{Gallagher2020} including \citet{Barklem00} damping.
After pre-computing a grid of NLTE curves of growth at many $\Teff$ and $\logg$ values, we use Delaunay triangulation and linear interpolation to find NLTE abundances as a function of stellar parameters and equivalent widths (using \texttt{scipy.interpolate.LinearNDInterpolator}).

In the rest of this paper, we adopt the NLTE [Ba/H] abundances as our fiducial abundances. However in Appendix~\ref{app:basys} we show a comparison between the MOOG LTE/ATLAS and MULTI NLTE/MARCS abundances. Overall, the NLTE effects are approximately $-0.3$ dex, resulting in lower abundances compared to LTE modeling.

For completeness, we investigated NLTE corrections for Mg \citep{Bergemann2017}, Ca \citep{Mashonkina2007}, and Fe \citep{Bergemann12,Mashonkina16}\footnote{Correction grids available at \url{https://nlte.mpia.de/} and \url{http://spectrum.inasan.ru/nLTE/}.}.
We found that [Mg/H] increased by $0.06 \pm 0.02$ dex and [Ca/H] increased by $0.12 \pm 0.04$ dex, where the uncertainty indicates the variation across the stellar parameter range.
For Fe, in most stars only the 6494.98{\AA} line is available to measure. Most NLTE correction grids do not include this line, with the exception of \citet{Mashonkina16} whose interpolation grid only extends up to 5000K. The correction for the three stars with $\Teff < 5000\unit{K}$ increases [Fe/H] by $0.15 \pm 0.01$ dex. 
We decided not to include these corrections in our results, though these mean offsets can be applied if desired.

\subsubsection{Abundance Summary}
In summary, the main abundance results are Ba, Fe, Ca, and Mg measurements or upper limits.
The VLT data are used to measure Ba, Fe, and Ca.
The M2FS high-resolution data are used to measure Mg and to verify Fe and Ca.
The M2FS medium-resolution data are used to verify Ba (Table~\ref{tab:abunds}).
For the brightest stars, we use the M2FS HiRes data to derive abundances for Cr, Ti, and Nd, which we report in Appendix~\ref{app:moreelems}. We do not discuss these elements more, as they are consistent with the discussion in \citetalias{Ji16c} and \citetalias{Roederer16b}.
We adopt NLTE abundances for Ba and LTE abundances for other elements.

\section{Results}\label{sec:results}

\subsection{Radial Velocity Distribution}
\begin{figure*}
    \centering
    \includegraphics[width=0.8\linewidth]{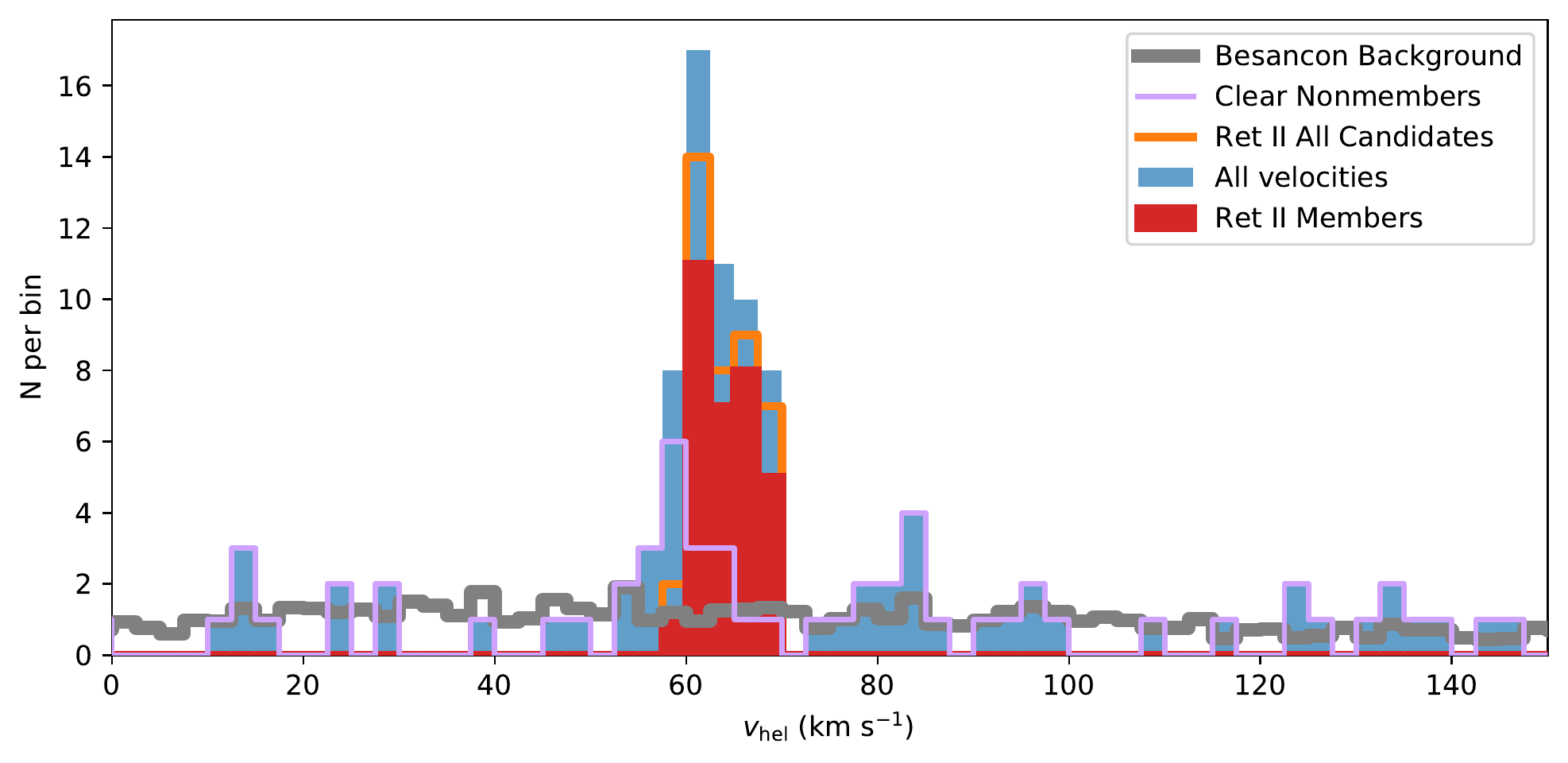}
    \caption{Histogram of measured velocities (blue), with definite (candidate) Ret~II members in red (orange).
    For comparison, velocities from the Besan{\c c}on model are shown as a grey histogram, normalized to match the observed sky area of 490 sq arcmin.
    The solid (dotted) purple histograms indicates subtracting the definite, red (candidate, orange) members from the full blue histogram.
    The non-members are overall consistent with the Besan{\c c}on background model, though there is still a small excess above background near the velocity of Ret II.
    }
    \label{fig:vhelhist}
\end{figure*}

Figure~\ref{fig:vhelhist} shows the radial velocities of all stars in our sample (solid blue histogram).
There is a clear peak at ${\approx}65$\,km\,s$^{-1}$ associated with Ret~II.
Stars considered clear Ret~II members are shown as the solid red histogram, while all Ret~II members including candidates are the open orange histogram (see Section~\ref{sec:membership} for details).
For comparison, we show the radial velocity distribution of a smooth background halo from the Besan{\c c}on model \citep{Besancon}.
We restrict the background to a CMD region surrounding our targets defined by four points $(g-r, g) = (0.2, 21.2), (0.7, 21.2), (0.45, 16.0), (1.0, 16.0)$. We query a large area for statistics and rescale the distribution to the FLAMES field of view.

The residual background after removing the clear and candidate members is shown as a purple histogram.
There continues to be an excess of stars near the velocity of Ret~II relative to the Besan{\c c}on model.
Detailed investigation of the proper motions and spatial position shows that these residual stars cannot be members of Ret II. There may potentially be additional structure in this region of the sky that is not part of Ret~II, though we do not see any clear spatial or proper motion trends.

We determine the mean velocity $\bar{v}$ and velocity dispersion $\sigma_v$ of Ret~II with a Gaussian scatter model.
Each star is assumed to have a true velocity distributed according to a Gaussian with mean and standard deviation $\bar{v}$ and $\sigma_v$, and the actual observed velocity has a Gaussian noise added to it with the individual velocity uncertainty in Table~\ref{tab:obs}.
The prior on the mean velocity is uniform with no bounds, and the prior on the scatter is uniform in log space from $\sigma_v \in [0.1, 10]$ \kms.
The posterior is sampled using \texttt{Stan} (\citealt{Stan}, following \citealt{Casey2021}).
All five likely binary stars are removed.
Using only the 27 clear likely single members of Ret~II, we obtain $\bar{v} = 63.9 \pm 0.5 \kmsec$ and $\sigma_v = 2.97^{+0.43}_{-0.35}\kmsec$.
Adding the 8 candidate members gives $\bar{v} = 64.0 \pm 0.5 \kmsec$ and $\sigma_v = 2.96^{+0.44}_{-0.36}\kmsec$, which is the same within the uncertainties.
Our velocity dispersion is consistent both with previous measurements of ${\approx}3.3\kmsec$ (\citetalias{Simon15,Walker15,Koposov15b}) and with the lower value of $2.8^{+0.7}_{-1.2}$ inferred by \citet{Minor19} who statistically accounted for binaries. Our velocity dispersion is about two times more precise given more stars and additional independent velocity measurements,
but we caution that our velocity dispersion uncertainties may be overly optimistic given the simple treatment of possible systematics (see Appendix~\ref{app:rv}).

\begin{deluxetable*}{lrr}
    \tablecolumns{3}
    \tablecaption{\label{tab:scatters}Reticulum~II Properties}
    \tablehead{Quantity & Value & Reference/Prior}
    \startdata
    RA (J2000) & 03:35:47.83 & \citet{MutluPakdil18} \\
    Dec (J2000) & $-$54:02:47.8 & \citet{MutluPakdil18} \\
    Position Angle (deg) & 68 $\pm 2$ & \citet{MutluPakdil18} \\
    Ellipticity & 0.6 $\pm 0.1$ & \citet{MutluPakdil18} \\
    Half-light radius (arcmin) & 6.3 $\pm 0.4$ & \citet{MutluPakdil18} \\
    Half-light radius (pc) & 58 $\pm 4$ & \citet{MutluPakdil18} \\
    Distance modulus & 17.5 $\pm 0.1$ & \citet{MutluPakdil18} \\
    Distance (kpc) & 31.4 $\pm 1.4$ & \citet{MutluPakdil18} \\
    \hline
    Heliocentric Radial Velocity (\kms) & $63.9 \pm 0.5$ & Uniform \\
    Velocity dispersion (\kms) & $2.97^{+0.43}_{-0.35}$ & $\log\sigma_v \sim \unifdistr{-1}{1}$ \\
    \hline
    Mean metallicity $\left\langle\mbox{[Fe/H]}\right\rangle$ & $-2.64 \pm 0.11$ & Uniform \\
    Metallicity dispersion $\sigma_{\rm [Fe/H]}$ & $0.32^{+0.10}_{-0.07}$ & $\log\sigma_{\mbox{[Fe/H]}} \sim \unifdistr{-2}{0}$ \\
    \hline
    Mean NLTE barium abundance $\left\langle\mbox{[Ba/H]}\right\rangle$ & $-1.68 \pm 0.07$ &  Uniform \\
    Barium dispersion $\sigma_{\rm [Ba/H]}$  & $0.05^{+0.08}_{-0.03}$ or $<0.20$ & $\log\sigma_{\mbox{[Ba/H]}} \sim \unifdistr{-2}{0}$ \\
    Fraction of $r$-enhanced stars & $0.72^{+0.10}_{-0.12}$ & $f_r = \unifdistr{0}{1}$ \\
    \hline
    Absolute Magnitude $M_{V}$ & $-3.1 \pm 0.1$ & \citet{MutluPakdil18} \\
    Stellar Mass $M_{\star}$ ($M_\odot$) & $10^{3.51 \pm 0.04}$ & Assuming $M/L=2.2$ \\
    Dynamical Mass $M_{\rm dyn, 1/2}$ ($M_\odot$) & $10^{5.6 \pm 0.2}$ & Using \citet{Wolf2010} \\
    \enddata
    \tablecomments{Rows where the third column has a prior are measured from this work.}
\end{deluxetable*}

\subsection{Abundance Trends}
\subsubsection{[Ba/H] vs [Fe/H]}
\begin{figure*}
    \centering
    \includegraphics[width=0.9\linewidth]{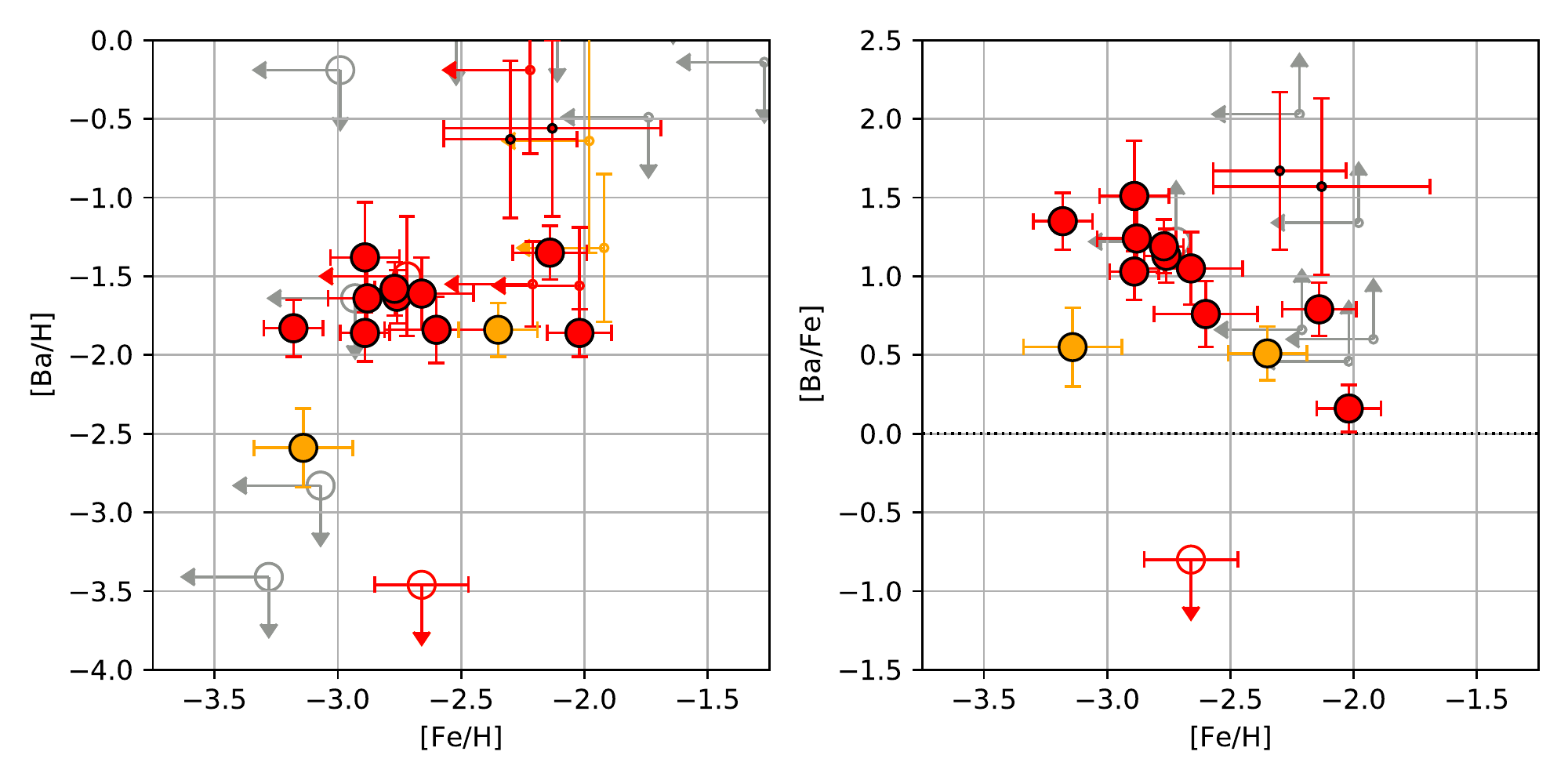}
    \caption{[Ba/H] and [Ba/Fe] vs [Fe/H] as measured by VLT/GIRAFFE.
    Red points indicate clear Ret II members and orange points indicate candidate Ret~II members.
    Large symbols are stars with SNR $> 25$, while small symbols are stars with SNR $<25$.
    $1\sigma$ error bars are shown for all detected abundances, and arrows drawn to indicate upper/lower limits.
    Open circles indicate stars with at least one upper limit, and stars are colored grey if they have no detection of either Ba or Fe.
    Considering only the clear members with high SNR (large red points), there is no apparent scatter in Ba but significant spread in Fe.
    }
    \label{fig:bafe}
\end{figure*}

Figure~\ref{fig:bafe} shows the [Fe/H], [Ba/H], and [Ba/Fe] abundances of our sample.
The red and orange points are clear and candidate Ret II members, respectively. Stars with limits on both [Fe/H] and [Ba/H] are shown in grey. Stars with SNR $>25$ are shown as larger points, while stars with SNR $<25$ are shown as smaller points. We only consider the large, high-SNR data points for the interpretation, but we show all the data for completeness.

There are two high SNR candidate member stars labeled on Fig~\ref{fig:bafe}. Star 97 is deemed a candidate member because it is extremely red compared to the fiducial Ret~II CMD. Its low inferred metallicity of $\mbox{[Fe/H]}=-2.4$ is inconsistent with its color unless it is an extremely carbon-enhanced star \citep[e.g.,][]{Koposov2018}. If it is a carbon-enhanced member, our model atmosphere grid may not be sufficient to accurately describe its properties. With this caveat in mind, the [Ba/H] for this star is right in line with most other Ret~II stars.
Star 14 is also deemed a candidate member due to its CMD position, which overlaps with a $\mbox{[Fe/H]}=-1.5$ isochrone. However, its weak Fe line suggests $\mbox{[Fe/H]} \sim -3.1$. This clear inconsistency suggests it is either a contaminant or its spectrum cannot be modeled using photometric stellar parameters. We thus do not strongly consider either candidate member's Ba abundance.

It is clear that the majority (${\approx}$2/3) of high-SNR Ret II stars with meaningful Ba abundance measurements lie in a constant [Ba/H] plateau, while the minority (${\approx}$1/3) have low undetected Ba abundances.
This corroborates the conclusions of \citet{Ji16b}, \citetalias{Ji16c}, and \citetalias{Roederer16b} that Ret~II is enriched by a single prolific $r$-process event, but now with more than two times the number of $r$-process abundance measurements or upper limits.
We note that for stars in the [Ba/H] plateau, there is a larger span in [Fe/H] than [Ba/H]. This is discussed in Section~\ref{sec:discussion}.

\subsubsection{[$\alpha$/Fe] vs [Fe/H]}
\begin{figure*}
    \centering
    \includegraphics[width=\linewidth]{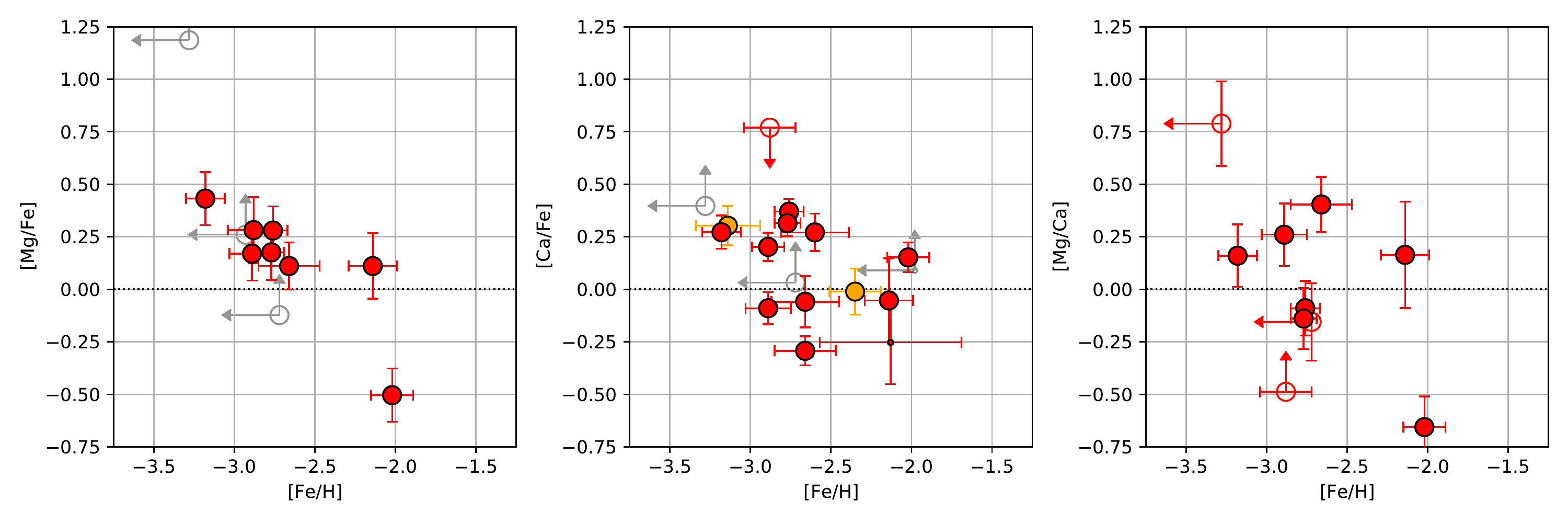}
    \caption{$\alpha$-element abundances in Ret~II.
    Symbols are same as Figure~\ref{fig:bafe}, with large red solid points indicating abundances of confident members, large orange points indicating abundances of candidate members, open red points indicating stars with one upper limit, and open grey points indicating stars with two upper limits.
    Note that two stars have Mg and Ca measurements but only Fe upper limits.
    \textit{Left:} [Mg/Fe] vs [Fe/H].
    \textit{Middle:} [Ca/Fe] vs [Fe/H].
    \textit{Right:} [Mg/Ca] vs [Fe/H].}
    \label{fig:abundmgca}
\end{figure*}

We measure two $\alpha$-elements, Mg and Ca, in several stars. 
The [$\alpha$/Fe] vs [Fe/H] results are shown in Figure~\ref{fig:abundmgca}. The number of stars is low and the error bars are large, but the left two panels show that both [Mg/Fe] and [Ca/Fe] broadly decline with increasing [Fe/H].
Given the low statistics, there is no clear ``$\alpha$-knee'' as seen in the Milky Way and some dwarf galaxies \citep{Venn04,Tolstoy09,Hill19}, though we note that most dwarf galaxies do not have such a clear knee \citep{Kirby20,Theler20}.
Regardless, both [Mg/Fe] and [Ca/Fe] show clear decreases somewhere between $-3.0 < \mbox{[Fe/H]} < -2.5$, a lower metallicity than the more massive classical dSph galaxies, as expected in a standard time-delay scenario \citep[e.g.,][]{Frebel12,Vargas13}.

While often assumed to vary together, in fact Mg and Ca can have different origins, since Mg is hydrostatically synthesized primarily in the most massive core-collapse supernova (CCSN) progenitors while Ca is explosively synthesized in most CCSN progenitors as well as Type Ia SNe \citep[e.g.,][]{McWilliam13,Hasselquist17,Ji20a}.
Most stars in Ret~II have similar Mg and Ca abundances, with two notable exceptions.
First, Star 5 has [Fe/H] $\sim -2.0$ and very low [Mg/Fe] (as previously pointed out by \citetalias{Ji16c}).
Second, Star 10 has [Fe/H] $< -3$ but a very high [Mg/Ca] $= +0.8$.
These two stars drive an overall decreasing slope in [Mg/Ca] with respect to [Fe/H], which could (but does not have to) imply an early excess (and late-time lack) of the most massive CCSNe in Ret~II.
\citet{Ji20a} pointed out that a decreasing [Mg/Ca] vs [Fe/H] trend is seen in several UFDs, but so far the trend is confined to UFDs that are kinematically associated with the Large Magellanic Cloud. An exciting possibility is that this indicates environment-dependent stellar populations or star formation histories. However, Ret~II is likely only recently captured by the Large Magellanic Cloud (\citealt{Patel20,Erkal20}, though see \citealt{Battaglia2022}).

Within the uncertainties, our Ca, Mg, and Fe measurements are consistent with previous results by \citetalias{Ji16c} and \citetalias{Roederer16b}. However, the lower SNR data in \citetalias{Ji16c} suggested a broadly flat [Ca/Fe] trend, while our more precise and larger number of Ca measurements now clearly suggest an overall decreasing [Ca/Fe] trend with [Fe/H].

\subsection{Intrinsic Scatter in [Fe/H]} \label{sec:fescat}

We measure the intrinsic [Fe/H] scatter using the 13 definite member stars with [Fe/H] detections.
The [Fe/H] distribution is modeled with a Gaussian \citep[e.g.,][]{Li18Carina}, and sampled using \texttt{Stan} in a manner similar to the velocity dispersion. We adopt a log-uniform prior for the intrinsic scatter of $10^{-2}-10^0$.
The results are a mean $\left<\mbox{[Fe/H]}\right> = -2.64$ with dispersion 0.32 dex, matching previous results \citep{Simon15,Walker15,Koposov15b}.
Adding Fe measurements from the two candidate members or including upper limits does not substantially affect these results.

\begin{figure*}
    \centering
    \includegraphics[width=\linewidth]{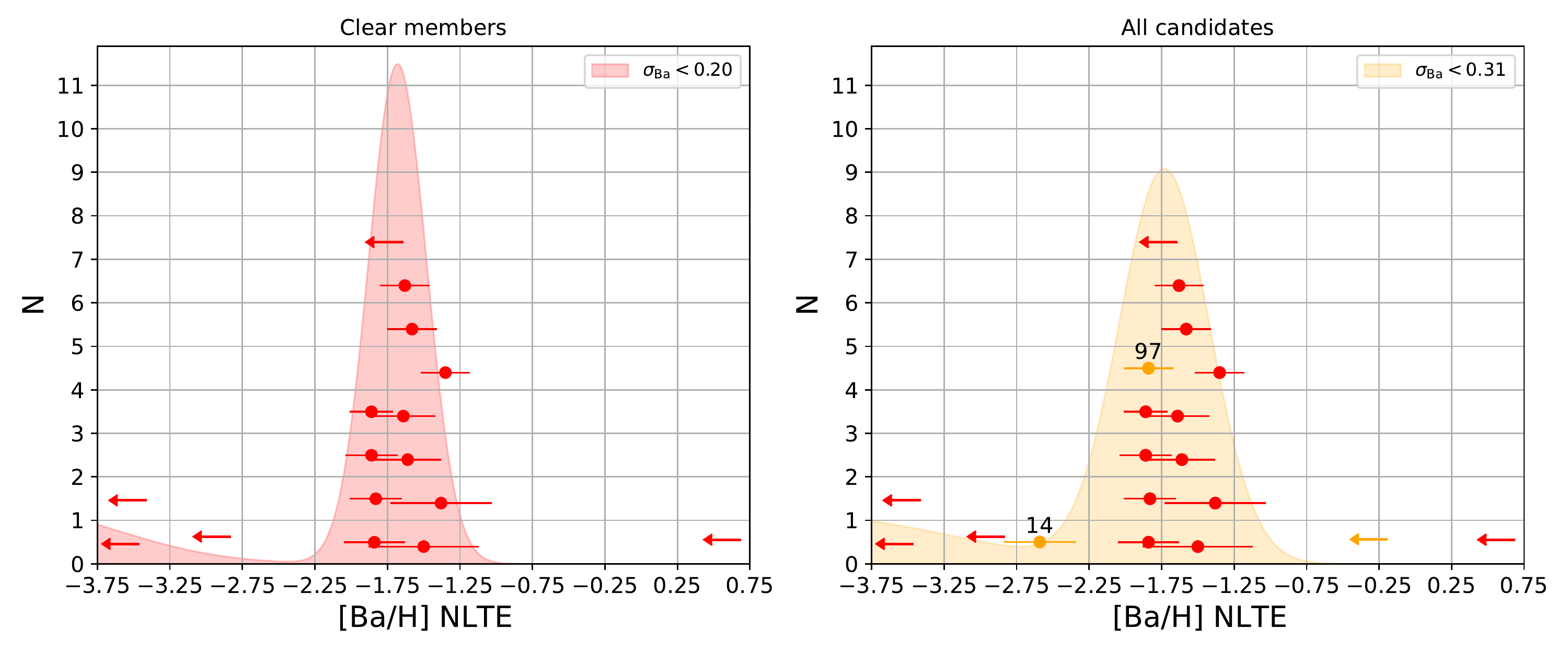}
    \caption{
    Barium measurements/upper limits and best-fit models for stars with SNR $>25$.
    The left panel shows our fiducial result only including clear members (16 stars), while the right panel also includes candidate members (3 more stars).
    Detections are shown as points with $1\sigma$ error bars, stacked within bins of 0.5 dex like a histogram. Stars with larger uncertainties are placed towards the bottom of each bin.
    Upper limits are shown as leftward-pointing arrows. Red and orange symbols indicate definite and candidate members, respectively.
    The shaded regions show the best-fit two-Gaussian model normalized to the number of stars on each panel.
    The width of both distributions ($\sigma_{\rm Ba}$) is unresolved, so we plot the width using the 95\% upper limit.
    Note that the model including candidates is wider primarily to accommodate Star 14.
    }
    \label{fig:abundhist}
\end{figure*}

\subsection{Intrinsic Scatter in [Ba/H]} \label{sec:bascat}

We detected [Ba/H] in 16 of our 32 member stars (21 out of 40 stars including candidate members).
However, only [Ba/H] measurements from stars with SNR $> 25$ were considered reliable due to sky subtraction residuals (see Figure~\ref{fig:vltfit} and Appendices~\ref{app:vtba} and \ref{app:basys}), restricting the [Ba/H] measurements to 11 out of 16 member stars (13 out of 19 candidate members).

Figure~\ref{fig:bafe} shows that the [Ba/H] measurements in Ret II have a clear peak at [Ba/H] $\sim -1.7$ and another group of stars with upper limits of [Ba/H] $\lesssim -3$ (as previously seen in \citealt{Ji16b,Roederer16b}\footnote{Star 7 has a much more stringent [Ba/H] upper limit in \citet{Ji16b}, but we kept its high [Ba/H] limit here to remain self-consistent.}).
We modeled this [Ba/H] distribution as a two-component Gaussian mixture.
The model has 5 parameters: two means, two intrinsic dispersions, and a mixing fraction. We assume a log-flat prior for the intrinsic dispersions from $10^{-2}-10^{0}$, a flat prior from 0 to 1 for the mixing fraction, and no constraint on the means (except that one has to be larger than the other).
For the abundance likelihoods, for stars with [Ba/H] detections we used the usual Gaussian likelihood using the [Ba/H] measurement and uncertainty as the mean and standard deviation for that star.
For the stars with [Ba/H] 99\% upper limits, we adopted a step-function likelihood where 99\% of the probability is uniform between [Ba/H]$=-5$ to the measured upper limit, and 1\% of the probability is uniform from the measured upper limit to [Ba/H]$=+1$.
This model was implemented in \texttt{Stan} 
and sampled using the NUTS sampler with 4 chains and $10^6$ steps.
Model parameters were initialized near likely values from initial visual examination (i.e.,\ mixing fraction based on the number of limits vs measurements, intrinsic scatters of 0.1 dex, component means of $-4.0$ and $-1.5$).
Our final chains were visually well-mixed and had $>10000$ effective samples for all parameters.

The results of our fiducial [Ba/H] dispersion fits are shown in Figure~\ref{fig:abundhist}.
Only stars with SNR $> 25$ are shown.
The points show the exact [Ba/H] values and their uncertainties, and they are stacked in the vertical direction in bins of 0.5 dex like a histogram.
As before, clear members are shown in red and candidate members in orange.
The arrows indicate the 99\% upper limits for stars with undetected Ba lines, which are included in our modeling procedure.
The best-fit two-component models are shown as shaded regions (red for clear members, orange for candidate members).

Overall, the detected Ba abundances for the clear member stars are clearly unimodal, and the per-star [Ba/H] uncertainty can explain the observed scatter in [Ba/H] abundances. We obtain a 95\% upper limit $\sigma_{\rm [Ba/H]} < 0.20$ dex.
If adding the candidate members, we obtain a weaker upper limit $\sigma_{\rm [Ba/H]} < 0.31$ dex, which is primarily driven by star 14 that has a weak Ba line. As discussed previously, if this star is a member it is not likely that its stellar parameters can be determined with photometry, so the [Ba/H] inference for this star is likely inaccurate.

As a note of caution, we believe the choices made for our fiducial measurement are the most appropriate for quantifying the [Ba/H] intrinsic scatter in Ret II, as they minimize the impact from known systematic issues.
But for completeness, in Appendix~\ref{app:basys} we show the effect of all choices (i.e.,\ SNR cut, membership, $\nu_t-\logg$ relation) on the intrinsic scatter measurement.
Changing the $\logg-\nu_t$ relation changes the limit to $\sigma_{\rm [Ba/H]} < 0.28$ dex for clear members and $\sigma_{\rm [Ba/H]} < 0.37$ dex for candidate members. This effect is driven almost entirely by the two cool stars (ID 1 and 97), which have low statistical uncertainty due to their brightness but have a strong dependence on microturbulence.

\subsection{Abundance Trends with Radius}
\begin{figure}
    \centering
    \includegraphics[width=0.9\linewidth]{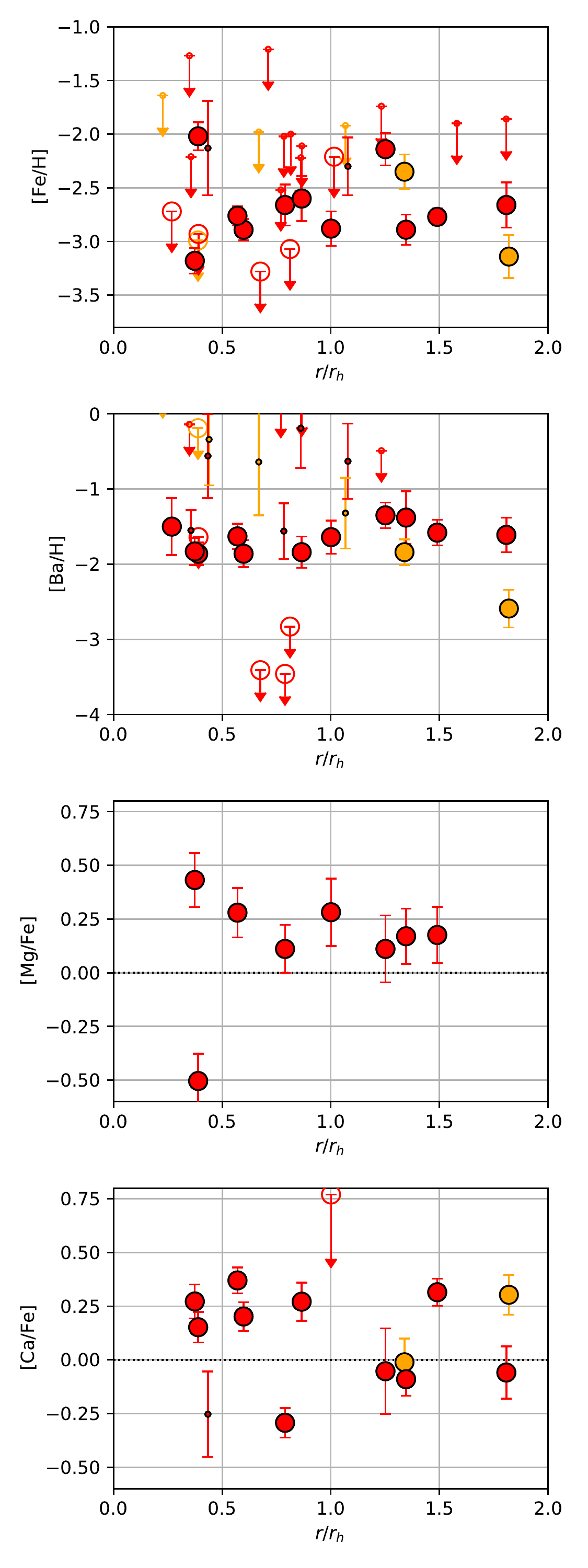}
    \caption{Abundance gradients. The x-axis on all plots is the elliptical radius, in units of the elliptical half-light radius.
    The symbols are as in Figure~\ref{fig:bafe}, with large red points indicating high-SNR stars with confident membership.
    There are no significant abundance gradients.}
    \label{fig:abundradius}
\end{figure}

Figure~\ref{fig:abundradius} shows trends in [Fe/H], [Ba/H], [Mg/Fe], and [Ca/Fe] with radius for confirmed and candidate members.
Overall, there are no strong abundance gradients for Fe and Ba.
This lack of trend within this radius is expected, as any initial abundance gradients produced when Ret~II was forming its stars at $z \gtrsim 6$ will likely be dynamically mixed away by $z=0$: for $10^4$ stars within the half-light radius of 58 pc and typical velocity of 3 \kms, the two-body relaxation time is about 2.5 Gyr \citep{BinneyTremaine}. Note a metallicity gradient could be present at larger radii, as illustrated by the clear gradient in Tucana II \citep{Chiti2021}, but our observations only reach two half-light radii.

For Mg and Ca, there may be a small abundance gradient in [Mg/Fe] and [Ca/Fe], perhaps decreasing slightly at larger distances.
For Mg, this could easily be due to small number statistics rather than an abundance gradient; and for Ca, the scatter around a mean trend would be comparable to or larger than the size of the gradient itself.
Thus, we do not consider there to be evidence for any abundance gradients out to two half-light radii.

\section{Discussion}\label{sec:discussion}

We have carefully determined the [Ba/H] distribution in Ret~II.
Figures~\ref{fig:bafe}~and~\ref{fig:abundhist} and Table~\ref{tab:scatters} show that ${\sim}30\%$ of the stars in Ret~II are relatively metal-poor with no detected $r$-process elements, while ${\sim}70\%$ of the stars in Ret~II have identical $r$-process abundances of $\mbox{[Ba/H]} = -1.7$, precise to 0.2 dex.
We now consider the implications of this measurement for galaxy formation and metal mixing (Section~\ref{disc:galmix}), the $r$-process site (Section~\ref{disc:rprocsite}), and chemical evolution in Ret~II (Section~\ref{disc:chemevol}). 
In most of the discussion we assume that the high [Ba/H] plateau in Ret~II originates predominantly from a single $r$-process event. However, we consider (and dismiss) the possibility of Ba contamination from multiple $r$-process sources or non-$r$-process sources in Section~\ref{disc:contamination}.

Previous studies indicated that the $r$-process material in Ret~II likely originates from a single $r$-process event, rather than multiple $r$-process events occurring within Ret~II \citep{Ji16b,Roederer16b}.
To update those conclusions with more recent results, out of 20 UFDs with high-resolution chemical abundances measured, Ret~II has $r$-process abundances 2-3 orders of magnitude higher than the neutron-capture element abundances in other UFDs. It is thus extremely unlikely that Ret~II would be enriched by multiple prolific $r$-process events, while other UFDs have no similarly prolific events.
To estimate this probability, we ran Monte Carlo simulations assuming that the number of $r$-process events in each of 20 galaxies is distributed as a Poisson random variable. We assume each UFD is equally likely to host an $r$-process event and run simulations where the intrinsic rate of $r$-process events is between 0.00 to 0.50 events per galaxy. We then conservatively take the highest probability.
The probability that any one galaxy gets $\geq 2$ $r$-process events while every other galaxy gets 0 $r$-process events is $<1.5\%$.
Two other ultra-faint satellites also appear to display a lower level of $r$-process enhancement (Tucana~III and Grus~II; \citealt{Hansen17,Hansen2020,Marshall18}).
If we assume that both of these systems are dwarf galaxies and hosted an $r$-process event, the probability that three UFDs have at least one $r$-process event and one has at least two $r$-process events is $<6\%$.  However, the classification of Tucana~III and Grus~II as dwarfs is currently uncertain \citep{Simon17,Simon2020}; if neither is a dwarf galaxy then the probability of multiple $r$-process events in Ret~II is again $\lesssim1.5\%$.
Thus, it is unlikely that Ret~II was enriched by more than one $r$-process event.

\subsection{Well-mixed [Ba/H] Implies Bursty Star Formation in Ret~II}\label{disc:galmix}

We first interpret the distribution of detected [Ba/H], which has a low intrinsic dispersion $\sigma_{\rm Ba} < 0.20$.
Because all this $r$-process material is deposited at a single time in Ret~II, the observed variation in [Ba/H] is entirely due to variations in metal mixing.
The unresolved Ba dispersion thus implies that the $r$-process material in Ret~II must have been very well-mixed by the time the high-Ba stars in Ret~II have formed.
To interpret this homogeneous Ba, consider a parcel of gas with a fresh source of metals.
There are two crucial timescales: $\tmix$, the time for the metals to completely mix within that gas, and $\tsf$, the time to turn that parcel of gas into stars.
Stars forming from this gas will be chemically homogeneous if mixing is faster than star formation, i.e. $\tmix < \tsf$.
The homogeneous Ba abundances in Ret~II thus provide a lower limit on the time between when the $r$-process event occurs and when the next generation of stars forms. In other words, the mixing timescale provides a constraint on the burstiness of star formation.
We argue in this section that both analytic estimates and simulations have shown that $\tsf > \tmix > 100 \text{~Myr}$, and thus there must be at least a 100 Myr gap in the star formation history of Ret~II. This is the first observational evidence of bursty star formation in an ultra-faint dwarf galaxy.

We first describe some basic physical ingredients determining $\tmix$.
Metal mixing in dwarf galaxies can be approximated as proceeding in two phases: the initial explosion remnant and subsequent turbulent mixing \citep[e.g.,][]{Karlsson2008,Emerick2020}.
For an individual explosion, the initial remnant is dominated by a momentum-driven snowplow, lasts only ${\sim}10^5$ yr, and sweeps up ${\sim}10^5 M_\odot$ of gas (for a $10^{51}$ erg explosion; e.g., \citealt{Cioffi88, Ryan96, Greif2007, Ji15, Macias2019, Magg2020}).
This mass is insignificant compared to a dwarf galaxy's ISM.
The dominant process determining $\tmix$ is thus turbulent mixing, where large-scale energy injection cascades to small-scale velocity eddies that enable microscopic diffusion \citep[e.g.,][]{Pan2013}.
Turbulent mixing is typically modeled as a diffusion process \citep{Klessen2003, Karlsson2008, Greif10, Ji15, Krumholz2018, Beniamini2020, Tarumi2020},
$R_t^2 \propto D_t \tau$,
where $R_t$ is the turbulent diffusion distance, $D_t$ is the turbulent diffusion coefficient, and $\tau$ is the time since material is deposited.
Drawing an analogy to mixing length theory, the diffusion coefficient can be estimated by a typical length and velocity scale driving turbulence, $D_t \sim R_{\rm turb} v_{\rm rms}$.
Complete mixing occurs when $R_t$ reaches a length scale associated with the full size of the galaxy, $R_t = R_{\rm gal}$.
Thus, $\tmix \sim R_{\rm gal}^2 / (R_{\rm turb} v_{\rm rms})$ \citep{Pan2013}.
In early dwarf galaxies, turbulence is primarily driven by gravitational gas accretion or mergers \citep{Wise2007,Greif2008,Klessen2010,SafranekShrader2012,Ritter15}, which can be used to estimate a turbulent diffusion coefficient \citep{Karlsson2008,Ji15}.
Putting in typical values for a dwarf galaxy ($R_{\rm gal} \sim 500$ pc the extent of the star forming gas, $R_{\rm turb} \sim 100$ pc the size of a typical explosion remnant, and $v_{\rm rms} \sim 10 \kms$ the turbulent velocity of ambient gas, from \citealt{Tarumi2020}) gives $\tmix \sim 250$ Myr.

While intuitive, the mixing length formalism fails to describe the full physics of the complex, anisotropic, and multi-phase metal-mixing process in dwarf galaxies. 
For example, it is well known that mixing depends on the temperature of the ISM phase that the metals reside in, such that hot ISM phases mix much more efficiently than cold ISM phases \citep[e.g.,][]{Kobulnicky1997,deAvillez2002,Emerick18,Emerick2019,Emerick2020}; and the anisotropic topology of cold gas clumps and filaments in early dwarf galaxies affects where metals get deposited and new stars form \citep[e.g.,][]{Webster16,Chiaki2018,Magg2022}.
It is thus crucial to study metal mixing with hydrodynamic galaxy formation simulations.
A few recent simulations have explicitly studied metal mixing in dwarf galaxies \citep[e.g.,][]{Webster14,Webster15,Hirai15,Hirai2017,Revaz16,Escala18,Emerick2019,Emerick2020,Tarumi2020,Jeon2021}.
In the vast majority of simulations, abundance scatter from individual metal sources is typically very large, ranging from 0.4-2.0 dex \citep[e.g.,][]{Safarzadeh17,Emerick2020,Applebaum2021}\footnote{Our scatter limit of 0.2 dex is a $1\sigma$ rms, for which almost no simulation provides a quantitative value. Since most UFDs in simulations are resolved by fewer than 100 star particles, we currently estimate the rms by taking the range of star particle abundances, which corresponds to $\pm 2\sigma$ interval, and dividing by 4. It would be helpful for future simulations to provide the actual rms values.}.
Note that many simulations compare their simulation abundance scatter directly to the observed abundance data, without deconvolving the abundance uncertainties.

\citet{Tarumi2020} performed a direct simulation of $r$-process enrichment in Ret~II, simulating a UFD and injecting $r$-process elements from a single NSM to see the resulting $r$-process abundance spread.
In order to homogenize the gas to a level consistent with that observed in Ret~II, they found that the gas had to mix for a time period of a few hundred Myr before forming stars.
They measure an effective diffusion coefficient in their simulation of $D_t \approx 10^{-3}$ kpc$^2$ Myr$^{-1}$, resulting in a complete mixing timescale of ${\approx}250$ Myr.
Thus, we expect $100 \text{~Myr} < \tmix < \tsf$, and the timescale between bursts of star formation in Ret~II should be over 100 Myr, a significant fraction of a Hubble time at $z > 6$.

There are a few important caveats to the \citet{Tarumi2020} simulation. First, the mixing time can be affected by stochastic events.
When \citet{Tarumi2020} exploded the NSM at the outskirts of the galaxy rather than the center (to mimic a velocity kick, also see \citealt{Safarzadeh17,Safarzadeh2019a,Bonetti2019}), the lower diffusion coefficient resulted in less efficient mixing of the $r$-process elements.
\citet{Emerick2019,Emerick2020} also emphasize that the exact timing and location of $r$-process production relative to other stellar feedback sources that drive turbulence can both increase and decrease the mixing time. 
Second, the abundances in the \citet{Tarumi2020} simulation do not match Ret~II observations: these simulations produce a flat trend in [Ba/Fe] vs [Fe/H], whereas Figure~\ref{fig:bafe} shows a flat trend in [Ba/H] vs [Fe/H]. A correlation between Ba and Fe can occur if Ba and Fe are well-mixed relative to each other but with varying overall metallicity differences. The simulations by \citet{Tarumi2020} indeed have a gas-rich merger that helps homogenize Ba and Fe but causes a dispersion in [X/H] at fixed time (Y. Tarumi, private communication).
Finally, most galaxy formation simulations stay at relatively low resolutions, e.g. \citet{Tarumi2020} adopt the ISM model from the Auriga Project that uses an effective equation of state model below 0.1 \unit{cm^{-3}} \citep{Grand2017}. This may resolve mixing in the large-scale ISM, but it may not resolve small-scale inhomogeneous mixing \citep[e.g.,][]{Pan2013,Chiaki2018,Magg2022}.
\citet{Tarumi2020} is the only simulation directly comparable to Ret~II, but 
the overall timescale of $>100$ Myr to mix matches results from other simulations \citep{Hirai2017,Emerick2019} and estimates based on the mixing length scaling relation \citep{Karlsson05,Pan2013,Ji15}. Thus, the mixing timescale of $>100$ Myr seems robust.

In summary, the mixing time in UFDs is likely larger than 100 million years. The homogeneous $r$-process abundances in Ret~II thus indicate that at least two early bursts of star formation occurred in Ret~II, separated by at least a few hundred million years.
The first burst produced $r$-process elements that enriched stars born in the second burst (which should be relatively extended due to the presence of Type~Ia enrichment, see Section~\ref{disc:chemevol}).
Given that we find ${\approx}$70\% of Ret~II stars are $r$-process enhanced, a concrete prediction is that star formation histories of Ret~II with a precision of ${\lesssim}100$Myr should show 30\% of stars forming first, a gap of $>100$ Myr, and then the other 70\% of star formation.
Resolving bursty star formation on 100 Myr timescales is currently out of reach of direct star formation history measurements \citep[e.g.,][]{Brown14,Weisz14a}, but it qualitatively matches predictions from several UFD simulations \citep[e.g.,][]{Wheeler2019,Jeon17}. These simulations achieve burstiness due to strong clustered supernova feedback, which purges the galaxy of star-forming gas and requires a long recovery time for gas to cool and form the next generation of stars \citep[e.g.,][]{Jeon14}. The r-process elements can thus homogenize during this extended recovery time.

\subsection{A Prompt, High-Yield R-Process Site}\label{disc:rprocsite}

The [Ba/H] distribution in Ret~II introduces some new constraints on the $r$-process site's yield and delay time. The mean $\left\langle\mbox{[Ba/H]}\right\rangle = -1.68 \pm 0.07$ provides a constraint on the ratio
$M_{\rm Ba}/M_{\rm H} = 136.5 \times 10^{\mbox{[Ba/H]} - 9.82} \approx 10^{-9.36}$, where $136.5$ is the average atomic mass of Ba and $9.82 = 12.00 - 2.18$ accounts for the \citet{Asplund09} solar composition.
Assuming an $r$-process ratio of $\mbox{[Ba/Eu]}=-0.80$ in Ret~II \citep{Ji18}, this means $\left\langle\mbox{[Eu/H]}\right\rangle = -0.88$, or $M_{\rm Eu}/M_{\rm H} = 152.0 \times 10^{\mbox{[Eu/H]} - 11.48} \approx 10^{-10.18}$, where 152.0 is the average atomic mass of Eu and $11.48 = 12.00 - 0.52$ accounts for the solar composition.
The mass ratio $M_{\rm Eu}/M_r$ is $10^{-3.0}$, assuming $M_r$ is elements with $A \geq 80$ from the solar $r$-process pattern \citep{Sneden08, Cote2018, Ji19b}.
Thus, we find that $M_r/M_{\rm H} \approx 10^{-7.2 \pm 0.1}$, where $M_{\rm H}$ is an effective dilution mass of hydrogen.

Inferring the yield $M_r$ of the $r$-process site depends on what is assumed for $M_{\rm H}$.
\citet{Ji16b} previously argued that expected dilution masses range from $10^5-10^7 M_\odot$. The lower end is set by the initial explosion remnant \citep[also see simulations by][]{Magg2022}, and the upper end is set by the total available gas in a $10^8 M_\odot$ dark matter halo that is likely to host Ret~II at high redshift. This would correspond to $M_r \sim 10^{-2.2}-10^{-0.2} M_\odot$.
However, simulations show that the effective dilution masses are near the minimum of $10^5 M_\odot$ only when metals are inhomogeneously mixed \citep[e.g.,][]{Chiaki2018,Jeon2021,Magg2022}.
The homogeneity of $r$-process elements in Ret~II thus suggests that the metals have diluted into a larger hydrogen mass, which should be in the range $10^6-10^7 M_\odot$. The higher dilution mass would imply a higher $r$-process yield of $M_r \sim 10^{-1.2}-10^{-0.2} M_\odot$.

An alternate way to estimate the $r$-process yield is to count up the amount of $r$-process elements locked into stars, then apply a correction factor for how much $r$-process material would not be captured in stars. A large fraction of $r$-process can be lost to the intergalactic medium due to the low gravitational potential of early dwarf galaxies \citep[e.g.,][]{Beniamini17,Brauer2021}.
Both empirically and theoretically, only $\lesssim 10^{-2}$ of metals are retained in dwarf galaxies \citep{Dekel2003,Robertson2005,Kirby11outflow,McQuinn2015}.
We can estimate the total mass of $r$-process in Ret~II using its present-day stellar mass of ${\approx}3300M_\odot$ \citep{MutluPakdil18}. Assuming a hydrogen mass fraction of 0.75, the total $r$-process mass contained in Ret~II today is $10^{-3.8} M_\odot$.
Thus the expected yield of the $r$-process site should be $M_r \gtrsim 10^{-1.8} M_\odot$, consistent with our previous estimate $M_r \sim 10^{-1.2}-10^{-0.2} M_\odot$.
Note that the dilution masses described in \citet{Ji16b} implicitly include this metal loss to the IGM, as the higher effective dilution masses can be thought of as corresponding to lower metal retention (also see Fig 11 of \citealt{Magg2022}).

The homogeneous [Ba/H] also suggests that the $r$-process site has to be fairly prompt in Ret~II. \citet{Ji16a} originally argued that the recovery times in UFDs (i.e., the time for gas to recollapse to the center of a halo after feedback) were longer than $10-100$\,Myr \citep{Jeon14,Ji15,BlandHaw15}. This allowed a significant fraction of ordinary neutron star mergers with $10-100$\,Myr delay times to merge and enrich the gas before star formation. However, if the $r$-process material must then subsequently mix for an additional ${\gtrsim}100$\,Myr to homogenize, this puts strong pressure on the $r$-process site to be very prompt in order to mix fully before turning into stars.
Additionally, \citet{Simon2022} recently obtained a precise star formation history for Ret~II using \textit{Hubble} color-magnitude diagrams. Combined with the result from this paper that 70\% of Ret~II stars are $r$-process enhanced, they concluded that the $r$-process time delay in Ret~II must be shorter than 500 Myr.

Together, the higher $r$-process yield and more prompt $r$-process event implied by homogeneous $r$-process mixing slightly favor rare core-collapse supernovae over neutron star mergers as the source of $r$-process elements in Ret~II.
Our higher expected $r$-process yield of $10^{-1.2}-10^{-0.2} M_\odot$ is a better match to the $0.08-0.3 M_\odot$ of $r$-process produced in collapsar disk winds \citep{Fryer2006,Surman2006,Siegel19,Miller2020}, but also consistent with magnetorotationally driven jets ($10^{-2.5}-10^{-1.5} M_\odot$ of $r$-process, \citealt{Mosta2018}) or neutron star mergers ($10^{-3}-10^{-1} M_\odot$, \citealt{Wu16,Radice18}; note GW170817 had $r$-process mass ${\approx}10^{-1.5 \pm 0.3} M_\odot$, \citealt{Drout17,Kilpatrick17,Tanaka2017,Tanvir2017,Chornock2017}), or common envelope jet supernovae \citep[e.g.,][]{Grichener2019,Grichener2022}.
The fact that there is a few hundred Myr delay after $r$-process enrichment also favors core-collapse supernovae.
However, very prompt and high-yield neutron star mergers are still on the table \citep{Beniamini16,Beniamini2019,Safarzadeh2019b}.

\subsection{No Gas Accretion During Most Ret~II Star Formation}\label{disc:chemevol}

Figure~\ref{fig:bafe} shows that the [Ba/H] abundance of the $r$-process rich stars stays very flat over an extended range of [Fe/H]. This can be seen quantitatively by comparing the metallicity (Fe) dispersion of $0.32^{+0.10}_{-0.07}$ dex to the $r$-process (Ba) dispersion of $<0.20$ dex.
The simplest interpretation of the larger Fe dispersion is that the $r$-process stars formed over some extended period of time where Ret~II was able to self-enrich with iron from supernovae, as expected for a dwarf galaxy \citep{Willman12}.
The flat [Ba/H] abundance would then clearly indicate that there is no pristine gas accretion nor any significant $r$-process production during the last 70\% of Ret~II's stellar mass growth. If there were significant pristine gas accretion during this time, it would reduce [Ba/H] at high [Fe/H]\footnote{\citet{Tsujimoto17} used a similar feature in Draco to argue for discrete $r$-process events, but in this relatively luminous galaxy there is a degeneracy between the number of $r$-process enrichment events and the presence/lack of gas accretion.}.

This gas cutoff scenario also could explain the [Mg/Ca] trend seen in Figure~\ref{fig:abundmgca} through the integrated galactic initial mass function (IGIMF) \citep{Weidner13}.
In this model, a gas-poor galaxy is unable to create the densest and largest molecular clouds, introducing an effective upper mass limit to stars formed.
Since Mg is predominantly produced in the most massive core-collapse supernovae and Ca is produced in all supernovae, a restricted gas supply will result in lower [Mg/Ca] abundances relative to a fully sampled IMF \citep{McWilliam13,Ji20a,Lacchin2020}.
Thus, a lack of gas accretion could explain both the flat [Ba/H] and the declining [Mg/Ca] observed in Ret~II.
This observation should simplify chemical evolution models aiming to reproduce the $r$-process abundance trends of Ret~II \citep[e.g.,][]{Komiya2016,Ojima2018,Molero2021,Cavallo2021}.

A lack of gas accretion may indicate something about the broader formation environment of Ret~II.
In particular, it is expected that UFDs like Ret~II are ultimately quenched by reionization \citep[e.g.,][]{Bullock00,Benson02,Brown14,RodriguezWimberly2019}, but it is not yet clear whether reionization immediately removes cold gas from halos or just restricts gas inflow \citep[e.g.,][]{Okamoto2008,Weisz14b,Jeon17,Bose2018,Wheeler2019}.
Since ${>}70\%$ of Ret~II stars form in the absence of significant gas accretion, it may be that it formed all these stars after reionization.

In this vein, it is interesting to note that Ret~II is a satellite of the Large Magellanic Cloud \citep{Patel20,Erkal20,Battaglia2022}. \citet{Sacchi2021} tentatively find that the star formation histories of LMC UFD satellites (including Ret~II) take longer to complete the last 10\% of their star formation history compared to Milky Way UFD satellites. This could suggest that LMC UFD satellites like Ret~II were relatively isolated when they formed compared to Milky Way UFD satellites, and thus experienced delayed reionization.
If so, it would support the concept of patchy reionization at the smallest galactic scales \citep[e.g.,][]{Lunnan2012,Aubert2018}.
More recently, \citet{Simon2022} have determined the star formation history of Ret~II with additional HST photometry. They find that Ret~II likely took ${\sim}$2.5 Gyr to form all its stars, largely corroborating the arguments from chemical evolution here.

For completeness, we note that the above discussion has implicitly assumed that $\tmix$ and $\tsf$ are the same for both Fe and Ba.
One can imagine scenarios where the timing of stellar feedback causes an early source of Ba to be mixed more than a later source of Fe \citep[e.g.,][]{Ritter15,Schonrich2019}. In this case, stars at all metallicities would form simultaneously. This scenario seems unlikely for Ret~II given the coherent evolution of Mg and Ca with [Fe/H], but it provides motivation to obtain more precise star formation histories in Ret~II.

\subsection{Contamination of Ba by Other Sources}\label{disc:contamination}
We have interpreted our Ba measurements in Ret~II as tracing pure $r$-process, based on the pure $r$-process patterns found in \citet{Ji16b} and \citet{Roederer16b}.
However in principle there could be three possible contaminating sources of Ba that are empirically found in UFDs:
(1) a low-yield source of Ba observed in most UFDs, of unknown origin \citep{Frebel15,Roederer17,Ji19a}, but possibly attributed to $r$-process in neutrino driven winds (\citetalias{Ji16c}, \citealt{Simon19}) or s-process in rotating massive stars \citep{Frischknecht16,Limongi2018,Tarumi2021};
(2) late-time AGB enrichment in the ISM \citep{Frebel16,Ji20a}; or
(3) mass transfer of s-process Ba from a binary companion \citep{Frebel14}.

None of these possible contaminants will impact our conclusions.
The impact of the first two sources is much less than the $r$-process content of Ret~II, and it can be estimated by considering Ba abundances in other UFDs.
The low-yield Ba source produces typical $\mbox{[Ba/H]} \sim -4$ \citep{Ji19a}.
Ba in the ISM from AGB stars is not often seen in UFDs given their short star formation durations, but where it is seen it reaches $\mbox{[Ba/H]}_{\rm LTE} \sim -2.5$ \citep{Frebel16,Ji20a}.
In both cases, the amount of contamination is at most 1/10 of the Ba in Ret~II, too low to make a significant perturbation.
For the third source, AGB mass transfer tends to produce [Ba/Fe] $\sim +2$, much more Ba than is observed in these stars \citep[e.g.,][]{Frebel14,Hansen2016}.
Thus, we find it extremely unlikely that Ba is tracing anything other than the single $r$-process event in Ret~II.

However, we note that one candidate member, Star 97, is quite red both in DES and \Gaia photometry.
It is possible this is due to large amounts of carbon on the surface of the star, in which case it may have experienced mass transfer possibly including Ba. If so, this further justifies excluding star 97 from our main results.

\section{Conclusion}\label{sec:conclusion}

We have obtained multi-object spectroscopy of red giant branch members in the ultra-faint dwarf galaxy Reticulum II using VLT/GIRAFFE and Magellan/M2FS.
Our redetermination of the velocity and metallicity dispersion is consistent with past results, and we detect no significant spatial gradients in the element abundances.

Ret~II is of special interest due to its enrichment by a single $r$-process event.
Our primary new result is a quantitative measurement of the [Ba/H] distribution (Figure~\ref{fig:abundhist}, Table~\ref{tab:scatters}), which is a unique probe of gas dynamics and metal mixing within a faint, currently gas-free dwarf galaxy.
Approximately 30\% of Ret~II stars have no detected $r$-process material, while the other 70\% are enriched to a high enhancement.
We place an upper limit of $\sigma_{\rm [Ba/H]} < 0.20$ dex on the intrinsic [Ba/H] dispersion of the high-Ba stars, which implies that the initial $r$-process enrichment needs to turbulently mix and homogenize for at least 100 Myr before stars form.
This is the first direct evidence of bursty star formation in a UFD.
The long mixing time also favors an $r$-process site that is very prompt and produces a high $r$-process yield (${\gtrsim}10^{-1.5} M_\odot$). We thus slightly favor rare core-collapse supernovae as the source of $r$-process elements in this galaxy due to their higher $r$-process yield, though prompt high-yield neutron star mergers are allowed as well.

Examining the chemical evolution in Ret~II, we find an overall declining [$\alpha$/Fe] vs [Fe/H] pattern as expected in dwarf galaxies. Since [Ba/H] is flat over an extended [Fe/H] range, this suggests that Ret~II did not accrete significant gas during the last 70\% of its star formation. This is consistent with the observed declining [Mg/Ca] ratio if Ret~II was too gas-poor to form the most massive core-collapse supernovae. The chemical evolution of Ret~II thus suggests that it may have formed in an underdense environment, consistent with its status as a satellite of the Large Magellanic Cloud.

These constraints on UFD formation and the $r$-process site demonstrate the power of dwarf galaxy archaeology.
By finding stars in a common formation environment, it becomes possible to ask questions that could not be answered if these same stars were found individually scattered through the Milky Way. Reticulum~II is unusually nearby and thus currently accessible for this type of study, but as the next generation of extremely large telescopes comes online, it will become possible to extend similar techniques to study the chemistry of ultra-faint dwarf galaxies throughout the Milky Way \citep{Ji19Decadal}.

\acknowledgments

We thank Edward Olszewski, Meghin Spencer, and Matthew Walker for their help acquiring the M2FS data.
APJ thanks Andy Casey, Anirudh Chiti, Dan Kelson, Andy McWilliam, and Ting Li for many discussions about data reduction, processing, and analysis. APJ thanks Paz Beniamini, Alyson Brooks, Benoit Cote, Andrew Emerick, Evan Kirby, Mordecai Mac-Low, Brian O'Shea, and Yuta Tarumi for enlightening discussions about metal mixing and chemical evolution.
We also thank our anonymous referee for insightful comments that improved this paper.
Most of this study was done while APJ was supported by NASA through Hubble Fellowship grant HST-HF2-51393.001 awarded by the Space Telescope Science Institute, which is operated by the Association of Universities for Research in Astronomy, Inc., for NASA, under contract NAS5-26555.
APJ also acknowledges support from the U.S.\ National Science Foundation (NSF) grant AST~2206264, a Carnegie Fellowship, the Thacher Research Award in Astronomy, and MCT.
JDS acknowledges support from the NSF grant AST~1714873.
IUR acknowledges support from NSF grants PHY~14-30152 (Physics Frontier Center/JINA-CEE), AST~1613536, and AST~1815403/1815767, AST~2205847, and the NASA Astrophysics Data Analysis Program, grant 80NSSC21K0627.
AF acknowledges support from NSF grant AST-1716251. 
EM, MM, RSK, and MB acknowledge support by the German Research Foundation (DFG) via the Collaborative Research Center SFB 881 {\em The Milky Way System} (subprojects A1, A5, A10, B1, B2, B8). RSK furthermore thanks for support from the Heidelberg Cluster of Excellence EXC 2181 (Project-ID 390900948) {\em STRUCTURES: A unifying approach to emergent phenomena in the physical world, mathematics, and complex data} funded by the German Excellence Strategy, and from the European Research Council in the ERC synergy grant {\em ECOGAL – Understanding our Galactic ecosystem: From the disk of the Milky Way to the formation sites of stars and planets} (project ID 855130).
MB is supported through the Lise Meitner grant from the Max Planck Society. 
This project has received funding from the European Research Council (ERC) under the European Union’s Horizon 2020 research and innovation programme (Grant agreement No. 949173).

Based on observations collected at the European Southern Observatory under ESO programme 0100.B-0502(A).
This paper includes data gathered with the 6.5~meter Magellan Telescopes located at Las Campanas Observatory, Chile.

This research has made use of the SIMBAD database, operated at CDS, Strasbourg, France \citep{Simbad}.
This research has made use of NASA's Astrophysics Data System Bibliographic Services.
This work has made use of the VALD database, operated at Uppsala University, the Institute of Astronomy RAS in Moscow, and the University of Vienna.

This work has made use of data from the European Space Agency (ESA) mission
{\it Gaia} (\url{https://www.cosmos.esa.int/gaia}), processed by the {\it Gaia}
Data Processing and Analysis Consortium (DPAC,
\url{https://www.cosmos.esa.int/web/gaia/dpac/consortium}). Funding for the DPAC
has been provided by national institutions, in particular the institutions
participating in the {\it Gaia} Multilateral Agreement.

This research uses services or data provided by the Astro Data Lab at NSF's National Optical-Infrared Astronomy Research Laboratory. NOIRLab is operated by the Association of Universities for Research in Astronomy (AURA), Inc. under a cooperative agreement with the National Science Foundation.

This project used public archival data from the Dark Energy Survey (DES). Funding for the DES Projects has been provided by the U.S. Department of Energy, the U.S. National Science Foundation, the Ministry of Science and Education of Spain, the Science and Technology Facilities Council of the United Kingdom, the Higher Education Funding Council for England, the National Center for Supercomputing Applications at the University of Illinois at Urbana–Champaign, the Kavli Institute of Cosmological Physics at the University of Chicago, the Center for Cosmology and Astro-Particle Physics at the Ohio State University, the Mitchell Institute for Fundamental Physics and Astronomy at Texas A\&M University, Financiadora de Estudos e Projetos, Fundação Carlos Chagas Filho de Amparo à Pesquisa do Estado do Rio de Janeiro, Conselho Nacional de Desenvolvimento Científico e Tecnológico and the Ministério da Ciência, Tecnologia e Inovação, the Deutsche Forschungsgemeinschaft and the Collaborating Institutions in the Dark Energy Survey.

The Collaborating Institutions are Argonne National Laboratory, the University of California at Santa Cruz, the University of Cambridge, Centro de Investigaciones Enérgeticas, Medioambientales y Tecnológicas–Madrid, the University of Chicago, University College London, the DES-Brazil Consortium, the University of Edinburgh, the Eidgenössische Technische Hochschule (ETH) Zürich, Fermi National Accelerator Laboratory, the University of Illinois at Urbana-Champaign, the Institut de Ciències de l’Espai (IEEC/CSIC), the Institut de Física d’Altes Energies, Lawrence Berkeley National Laboratory, the Ludwig-Maximilians Universität München and the associated Excellence Cluster Universe, the University of Michigan, the National Optical Astronomy Observatory, the University of Nottingham, The Ohio State University, the OzDES Membership Consortium, the University of Pennsylvania, the University of Portsmouth, SLAC National Accelerator Laboratory, Stanford University, the University of Sussex, and Texas A\&M University.

Based in part on observations at Cerro Tololo Inter-American Observatory, National Optical Astronomy Observatory, which is operated by the Association of Universities for Research in Astronomy (AURA) under a cooperative agreement with the National Science Foundation.

\facilities{Magellan-Clay (M2FS), VLT-UT2 (FLAMES/GIRAFFE)}
\software{\texttt{MOOG} \citep{Sneden73,Sobeck11},
\texttt{smhr} \citep{Casey14,Ji20b},
\texttt{emcee} \citep{emcee},
\texttt{Stan} \citep{Stan},
\texttt{dynesty} \citep{dynesty},
\texttt{numpy} \citep{numpy}, 
\texttt{scipy} \citep{scipy}, 
\texttt{matplotlib} \citep{matplotlib},
\texttt{pandas} \citep{pandas},
\texttt{seaborn} \citep{seaborn},
and \texttt{astropy} \citep{astropy,astropy2}
}

\appendix

\section{Radial Velocities and Binarity} \label{app:rv}

As a relatively nearby UFD of great scientific interest, Ret~II has obtained many different epochs of radial velocities. We have collected all currently available literature velocities in Table~\ref{apptab:rv}.
The literature velocities are mostly derived from coadding spectra taken across 1-4 adjacent nights, so the MJD reported here is only accurate to 2 days of precision.
These velocities are not homogeneous and may suffer from systematic effects.

\begin{deluxetable}{l|cc|rrrrrrrr|rrrr|rrrr}
\tablecolumns{19}
\setlength{\tabcolsep}{3pt}
\tablecaption{\label{apptab:rv}Literature Radial Velocities}
\tablehead{
    ID &
    $v_{\rm hel}$ & $\sigma_{\rm vhel}$ & 
    $v_{\rm S15}$  & $\sigma_{\rm S15}$  & 
    $v_{\rm K15}$  & $\sigma_{\rm K15}$  & 
    $v_{\rm J16}$  & $\sigma_{\rm J16}$  & 
    $v_{\rm R16}$  & $\sigma_{\rm R16}$  & 
    $v_{\rm VLT}$  & $\sigma_{\rm VLT}$  & 
    $v_{\rm HR}$ & $\sigma_{\rm HR}$ & 
    $v_{\rm W15}$  & $\sigma_{\rm W15}$  & 
    $v_{\rm MR}$  & $\sigma_{\rm MR}$
\\
    \hline &&&
    \multicolumn{2}{c|}{MJD=57072}&
    \multicolumn{2}{c|}{MJD=57090}&
    \multicolumn{2}{c|}{MJD=57298}&
    \multicolumn{2}{c|}{MJD=57341}&
    \multicolumn{2}{c|}{MJD=58052}&
    \multicolumn{2}{c|}{MJD=58073}&
    \multicolumn{2}{c|}{MJD=57072}&
    \multicolumn{2}{c}{MJD=57639}
}
\startdata
  1 & $+65.3$ & $  0.2$ & \nodata & \nodata & $+66.3$ & $  0.2$ & $+66.8$ & $  1.0$ & $+65.5$ & $  1.0$ & $+67.6$ & $  0.7$ & \nodata & \nodata & \nodata & \nodata & $+56.3$ & $  2.4$ \\
  2 & $+61.1$ & $  3.2$ & $+59.1$ & $  0.9$ & $+61.4$ & $  0.4$ & $+62.7$ & $  1.0$ & $+62.0$ & $  1.0$ & $+65.5$ & $  0.7$ & $+64.9$ & $  0.7$ & $+61.8$ & $  0.4$ & $+55.1$ & $  2.4$ \\
  3 & $+62.0$ & $  0.4$ & $+62.3$ & $  1.0$ & \nodata & \nodata & $+62.0$ & $  1.0$ & $+62.2$ & $  1.0$ & $+63.5$ & $  0.7$ & $+66.8$ & $  0.8$ & $+63.2$ & $  0.5$ & $+52.4$ & $  2.4$ \\
  4 & $+58.5$ & $  0.3$ & $+57.7$ & $  1.0$ & $+59.6$ & $  0.5$ & $+60.9$ & $  1.0$ & $+59.7$ & $  1.0$ & $+61.2$ & $  0.7$ & $+60.4$ & $  0.8$ & $+60.4$ & $  0.7$ & $+47.3$ & $  2.4$ \\
  5 & $+61.7$ & $  0.3$ & \nodata & \nodata & $+63.5$ & $  0.5$ & $+61.9$ & $  1.0$ & \nodata & \nodata & $+63.8$ & $  0.7$ & $+63.2$ & $  0.8$ & \nodata & \nodata & $+54.0$ & $  2.4$ \\
  6 & $+64.3$ & $  0.4$ & $+64.4$ & $  1.1$ & $+65.6$ & $  0.9$ & $+63.5$ & $  1.0$ & \nodata & \nodata & $+67.4$ & $  0.7$ & $+67.0$ & $  0.8$ & $+62.9$ & $  1.2$ & $+59.7$ & $  2.4$ \\
  7 & $+63.2$ & $  0.4$ & $+65.2$ & $  1.2$ & $+65.9$ & $  1.2$ & $+62.7$ & $  1.0$ & \nodata & \nodata & $+65.7$ & $  0.7$ & $+65.0$ & $  0.9$ & $+63.9$ & $  2.3$ & $+43.6$ & $  2.4$ \\
  8 & $+60.2$ & $  0.4$ & $+59.8$ & $  1.2$ & $+61.9$ & $  0.8$ & $+61.9$ & $  1.0$ & \nodata & \nodata & $+62.5$ & $  0.7$ & $+61.7$ & $  0.8$ & $+61.8$ & $  1.4$ & $+51.6$ & $  2.4$ \\
  9 & $+69.2$ & $  0.4$ & $+69.7$ & $  1.4$ & $+70.8$ & $  1.1$ & $+71.6$ & $  1.0$ & \nodata & \nodata & $+71.2$ & $  0.7$ & $+70.8$ & $  0.8$ & $+70.0$ & $  1.7$ & $+59.6$ & $  2.4$ \\
 10 & $+62.1$ & $  3.9$ & $+62.3$ & $  1.1$ & $+69.1$ & $  1.0$ & \nodata & \nodata & \nodata & \nodata & $+64.0$ & $  0.7$ & $+61.4$ & $  0.8$ & $+65.6$ & $  1.1$ & \nodata & \nodata \\
 11 & $+67.0$ & $  0.9$ & $+67.9$ & $  1.1$ & $+65.4$ & $  1.8$ & \nodata & \nodata & \nodata & \nodata & \nodata & \nodata & $+70.2$ & $  3.2$ & $+67.9$ & $  1.3$ & $+61.3$ & $  2.4$ \\
 12 & $+64.6$ & $  0.5$ & $+65.7$ & $  1.1$ & $+65.0$ & $  1.4$ & \nodata & \nodata & \nodata & \nodata & $+67.2$ & $  0.7$ & $+66.7$ & $  0.8$ & $+69.1$ & $  1.5$ & $+53.2$ & $  2.4$ \\
 13 & $+63.6$ & $  2.6$ & $+65.6$ & $  1.3$ & $+68.2$ & $  1.7$ & \nodata & \nodata & \nodata & \nodata & $+67.8$ & $  0.7$ & $+63.1$ & $  0.8$ & $+70.4$ & $  1.9$ & \nodata & \nodata \\
 14 & $+62.3$ & $  0.7$ & $+59.3$ & $  1.8$ & \nodata & \nodata & \nodata & \nodata & \nodata & \nodata & $+65.3$ & $  0.7$ & $+69.0$ & $  3.3$ & \nodata & \nodata & \nodata & \nodata \\
 15 & $+62.6$ & $  0.6$ & $+63.2$ & $  1.4$ & $+63.4$ & $  1.7$ & \nodata & \nodata & \nodata & \nodata & $+65.3$ & $  0.7$ & $+65.1$ & $  3.3$ & $+62.5$ & $  1.9$ & \nodata & \nodata \\
 16 & $+64.6$ & $  0.9$ & $+59.1$ & $  8.2$ & \nodata & \nodata & \nodata & \nodata & \nodata & \nodata & $+67.4$ & $  0.9$ & $ +0.0$ & $  0.0$ & \nodata & \nodata & \nodata & \nodata \\
 17 & $+60.2$ & $  0.7$ & $+57.4$ & $  2.4$ & $+60.0$ & $  2.1$ & \nodata & \nodata & \nodata & \nodata & $+63.3$ & $  0.7$ & $+64.9$ & $  3.3$ & $+60.1$ & $  2.1$ & $+40.7$ & $  2.4$ \\
 18 & $+59.3$ & $  7.1$ & $+66.3$ & $  1.4$ & $+62.9$ & $  3.7$ & \nodata & \nodata & \nodata & \nodata & $+59.4$ & $  0.8$ & $+73.6$ & $  3.3$ & \nodata & \nodata & \nodata & \nodata \\
 19 & $+60.7$ & $ 10.5$ & $+67.9$ & $  1.4$ & $+78.9$ & $  1.8$ & \nodata & \nodata & \nodata & \nodata & $+57.9$ & $  0.8$ & $+64.4$ & $  3.3$ & $+70.2$ & $  3.3$ & $+26.5$ & $  2.4$ \\
 20 & $+63.6$ & $  0.7$ & $+63.5$ & $  1.4$ & \nodata & \nodata & \nodata & \nodata & \nodata & \nodata & $+66.3$ & $  0.7$ & $+70.9$ & $  3.4$ & $+66.4$ & $  2.9$ & \nodata & \nodata \\
 21 & $+61.0$ & $  0.8$ & $+56.7$ & $  1.9$ & \nodata & \nodata & \nodata & \nodata & \nodata & \nodata & $+64.7$ & $  0.9$ & $+64.2$ & $  3.4$ & \nodata & \nodata & \nodata & \nodata \\
 22 & $+65.6$ & $  0.7$ & $+64.7$ & $  1.8$ & \nodata & \nodata & \nodata & \nodata & \nodata & \nodata & $+68.6$ & $  0.8$ & $+68.4$ & $  3.5$ & $+66.7$ & $  2.0$ & \nodata & \nodata \\
 23 & $+61.6$ & $  0.7$ & $+59.8$ & $  1.8$ & \nodata & \nodata & \nodata & \nodata & \nodata & \nodata & $+64.7$ & $  0.8$ & $+65.2$ & $  3.5$ & \nodata & \nodata & \nodata & \nodata \\
 24 & $+63.6$ & $  0.8$ & $+68.0$ & $  3.5$ & \nodata & \nodata & \nodata & \nodata & \nodata & \nodata & $+65.9$ & $  0.8$ & $+71.6$ & $  3.4$ & \nodata & \nodata & \nodata & \nodata \\
 25 & $+62.0$ & $  0.7$ & $+61.9$ & $  2.0$ & \nodata & \nodata & \nodata & \nodata & \nodata & \nodata & $+64.9$ & $  0.8$ & $+63.0$ & $  3.4$ & $+65.0$ & $  2.9$ & \nodata & \nodata \\
 26 & $+61.5$ & $  0.8$ & $+61.7$ & $  4.8$ & \nodata & \nodata & \nodata & \nodata & \nodata & \nodata & $+64.0$ & $  0.9$ & $+69.1$ & $  4.0$ & \nodata & \nodata & \nodata & \nodata \\
 97 & $+67.5$ & $  0.7$ & \nodata & \nodata & \nodata & \nodata & \nodata & \nodata & \nodata & \nodata & $+70.3$ & $  0.7$ & \nodata & \nodata & \nodata & \nodata & \nodata & \nodata \\
 99 & $+67.2$ & $  0.8$ & \nodata & \nodata & \nodata & \nodata & \nodata & \nodata & \nodata & \nodata & $+69.9$ & $  0.8$ & $+70.9$ & $  3.8$ & \nodata & \nodata & \nodata & \nodata \\
100 & $+61.0$ & $  0.8$ & \nodata & \nodata & \nodata & \nodata & \nodata & \nodata & \nodata & \nodata & \nodata & \nodata & $+63.4$ & $  0.8$ & \nodata & \nodata & \nodata & \nodata \\
102 & $+66.8$ & $  0.7$ & \nodata & \nodata & \nodata & \nodata & \nodata & \nodata & \nodata & \nodata & \nodata & \nodata & $+69.1$ & $  0.7$ & \nodata & \nodata & \nodata & \nodata \\
134 & $+62.2$ & $  1.0$ & \nodata & \nodata & \nodata & \nodata & \nodata & \nodata & \nodata & \nodata & $+65.0$ & $  1.0$ & $ +0.0$ & $  0.0$ & \nodata & \nodata & \nodata & \nodata \\
142 & $+61.0$ & $  0.7$ & \nodata & \nodata & \nodata & \nodata & \nodata & \nodata & \nodata & \nodata & $+64.0$ & $  0.9$ & $+63.0$ & $  1.1$ & \nodata & \nodata & \nodata & \nodata \\
143 & $+58.9$ & $  0.9$ & \nodata & \nodata & \nodata & \nodata & \nodata & \nodata & \nodata & \nodata & $+61.7$ & $  0.9$ & $ +0.0$ & $  0.0$ & \nodata & \nodata & \nodata & \nodata \\
144 & $+65.0$ & $  1.1$ & \nodata & \nodata & \nodata & \nodata & \nodata & \nodata & \nodata & \nodata & $+67.6$ & $  1.2$ & $+68.5$ & $  2.6$ & \nodata & \nodata & \nodata & \nodata \\
151 & $+68.0$ & $  1.0$ & \nodata & \nodata & \nodata & \nodata & \nodata & \nodata & \nodata & \nodata & $+70.8$ & $  1.0$ & \nodata & \nodata & \nodata & \nodata & \nodata & \nodata \\
154 & $+61.9$ & $  1.0$ & \nodata & \nodata & \nodata & \nodata & \nodata & \nodata & \nodata & \nodata & $+64.6$ & $  1.0$ & \nodata & \nodata & \nodata & \nodata & \nodata & \nodata \\
157 & $+67.9$ & $  0.9$ & \nodata & \nodata & \nodata & \nodata & \nodata & \nodata & \nodata & \nodata & $+70.5$ & $  0.9$ & $+73.1$ & $  3.4$ & \nodata & \nodata & \nodata & \nodata \\
188 & $+68.4$ & $  1.2$ & \nodata & \nodata & \nodata & \nodata & \nodata & \nodata & \nodata & \nodata & $+71.2$ & $  1.2$ & \nodata & \nodata & \nodata & \nodata & \nodata & \nodata \\
192 & $+69.1$ & $  0.9$ & \nodata & \nodata & \nodata & \nodata & \nodata & \nodata & \nodata & \nodata & $+72.2$ & $  0.9$ & $+66.9$ & $  3.5$ & \nodata & \nodata & \nodata & \nodata \\
195 & $+68.1$ & $  0.8$ & \nodata & \nodata & \nodata & \nodata & \nodata & \nodata & \nodata & \nodata & $+71.1$ & $  0.8$ & $+65.2$ & $  4.2$ & \nodata & \nodata & \nodata & \nodata \\
\enddata
\end{deluxetable}

In a first attempt to calibrate the systematic effects, we adopt the \citetalias{Simon15} velocities as a reference velocity scale, as they have the most stars with common velocities compared to other literature sources.
For matched stars in each sample, we calculate a weighted mean velocity offset. After removing this offset, stars with velocity variations inconsistent with a chi-squared test with $p < 0.01$ are identified as likely binaries \citep[e.g.,][]{Chiti2022}.
We iterate this process until convergence, resulting in five binary stars: Star 2, 13, 18, 19, and 21.
The final mean velocity offsets relative to \citetalias{Simon15} are 1.01 \kms for \citetalias{Koposov15b}, 0.63 \kms for \citetalias{Ji16c}, 1.02 \kms for \citet{Roederer16b}, 2.79 \kms for our VLT spectra, and 2.37 \kms for our HighRes M2FS spectra.
Our MedRes M2FS data have a very large offset of $-8.9\kms$, and we thus decided to exclude it from any velocity studies.

\section{Effect of Microturbulence on Barium Abundances}\label{app:vtba}

Microturbulence ($\nu_t$) is a parameter introduced to 1D stellar atmospheres to account for unmodeled 3D atmospheric effects.
It affects lines at the saturated part of the curve of growth, where a higher microturbulence effectively desaturates strong lines by adding some extra doppler broadening \citep[e.g.,][]{GrayTextbook}.
When lots of Fe I lines are available, microturbulence is usually found by balancing Fe I abundance as a function of line strength.
Empirical measurements of microturbulence show that red giants with lower surface gravities (and temperatures) tend to have higher microturbulence values \citep[e.g.,][]{Barklem05}.

In our VLT spectra, the Ba 6496{\AA} line is at or near saturation when detected. For our coolest giants, increasing $\nu_t$ by 0.2 \kms reduces [Ba/H] by 0.15 dex.
Since we aim to resolve abundance scatter on the order of 0.20 dex, this systematic effect is often the dominant uncertainty, especially for the brightest giants where the equivalent width is well-measured.
There are not enough Fe lines to self-consistently measure $\nu_t$ in our stars, so we must use existing correlations between $\logg$ and $\nu_t$.
Here, we investigate five different datasets that measured $\nu_t$ using high-resolution spectroscopy of metal-poor red giant stars, examining systematic differences in $\nu_t-\logg$ relations, as well as the scatter around those relations.
The effect of these choices on our barium abundance results is investigated in Appendix~\ref{app:basys}.

Figure~\ref{fig:vtsys} shows the result of our investigation.
The left column in Figure~\ref{fig:vtsys} plots $\logg$ vs $\nu_t$ for data from \citet{Barklem05} (B05), \citet{Marino08} (M08), \citet{Cohen13} (C13), \citet{Roederer14c} (R14), and \citet{Jacobson15} (J15), while the other columns show the measured [Ba/H]$_{\rm LTE}$ compared to $\Teff$, [Fe/H], and the signal-to-noise ratio (SNR).
We separate \citet{Roederer14c} into a giants-only sample as well (R14g).
When available, we use the stellar parameters as tabulated in JINAbase \citep{jinabase}.
Horizontal branch stars have higher microturbulence than RGB stars, so they are removed with a cut $\logg > 0.00286\,\Teff - 12.7$.
To each dataset, we fit a linear and quadratic polynomial for $\nu_t$ as a function of $\logg$, using a robust fitter based on the routine \texttt{robust\_poly\_fit} in the AstroIDL library. The scatter around the fit is measured with the biweight scale of the residuals (a robust standard deviation).
The coefficients and scatter around each relation are given in Table~\ref{apptab:vtparams}.

Note that the right column of Figure~\ref{fig:vtsys} shows that our stars with SNR $< 25$ have a much larger scatter than stars above that threshold. Visual examination of the spectra (Figure~\ref{fig:vltfit}) suggests that these stars are adversely affected by inaccurate sky subtraction that is blended with the Ba line. We thus have decided to exclude the low-SNR stars from most analyses. Also note that we used MOOG LTE/ATLAS [Ba/H] abundances for this figure, though the conclusions are robust if using NLTE instead.

\subsection{Mean Relations}
There are clear differences in the average $\nu_t-\logg$ relation across different literature samples: at low $\logg$, the first three rows (B05, C13, J15) have systematically higher $\nu_t$ than the last three rows (M08, R14, R14g). 
The origin of this difference is not clear. One possibility is the spectra for M08 and R14 have SNR $\sim 100$, substantially higher than B05 and J15 which typically have SNR $\sim 30$. The low SNR of weak iron lines could bias microturbulence too high (as described by \citealt{Magain84}).
However, C13 also have SNR $\sim 100$ and obtain higher microturbulence values.
Another possibility is that NLTE effects bias the microturbulence due to underlying correlations between excitation potential, line strength, and typical NLTE correction size \citep{Bergemann12}.
Further investigation of these differences would be valuable but is beyond the scope of this paper appendix.

\begin{figure}
    \centering
    \includegraphics[width=0.80\linewidth]{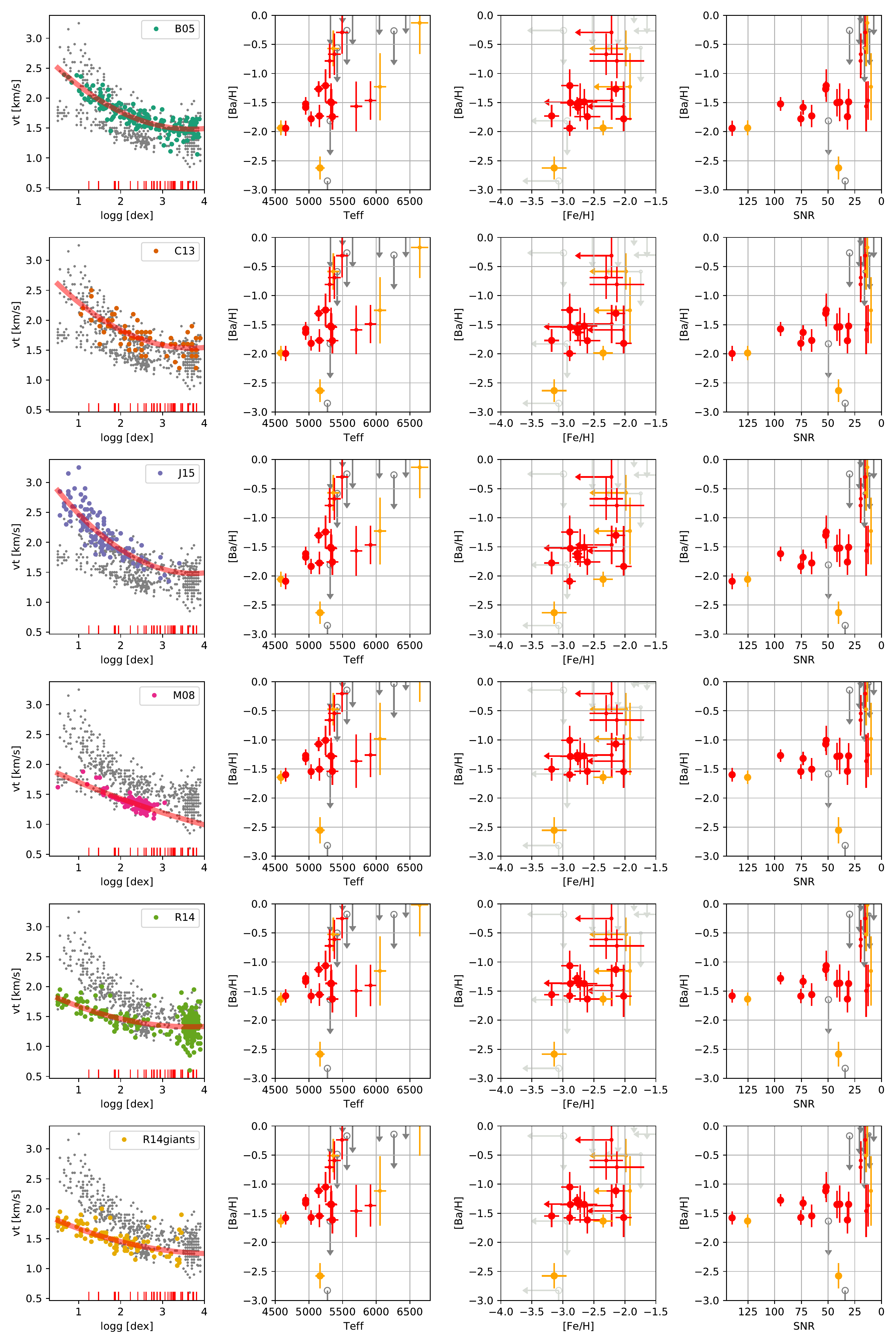}
    \caption{Impact of different $\nu_t-\logg$ relations.
    Left column: $\logg$ vs $\nu_t$ for each sample of stars. The thin red lines at the bottom indicate the $\logg$ of Ret~II stars. The best quadratic fit is plotted as a thick red line. The grey points show stars from all other rows for context. Left-middle column: [Ba/H] vs $\Teff$. Right-middle column: [Ba/H] vs [Fe/H]. Right column: [Ba/H] vs SNR. Stars with SNR $< 25$ display substantially larger [Ba/H] scatter, likely due to residuals from sky subtraction.
    Overall, the first three rows (B05, C13, J15) have similar trends, showing an upturn in $\nu_t$ at low $\logg$ but a very noticable trend in [Ba/H] vs $\Teff$.
    The last three rows have smaller $\nu_t$ and little trend in [Ba/H] vs $\Teff$ (excluding stars with SNR $< 25$).}
    \label{fig:vtsys}
\end{figure}

\begin{figure}
    \centering
    \includegraphics[width=\linewidth]{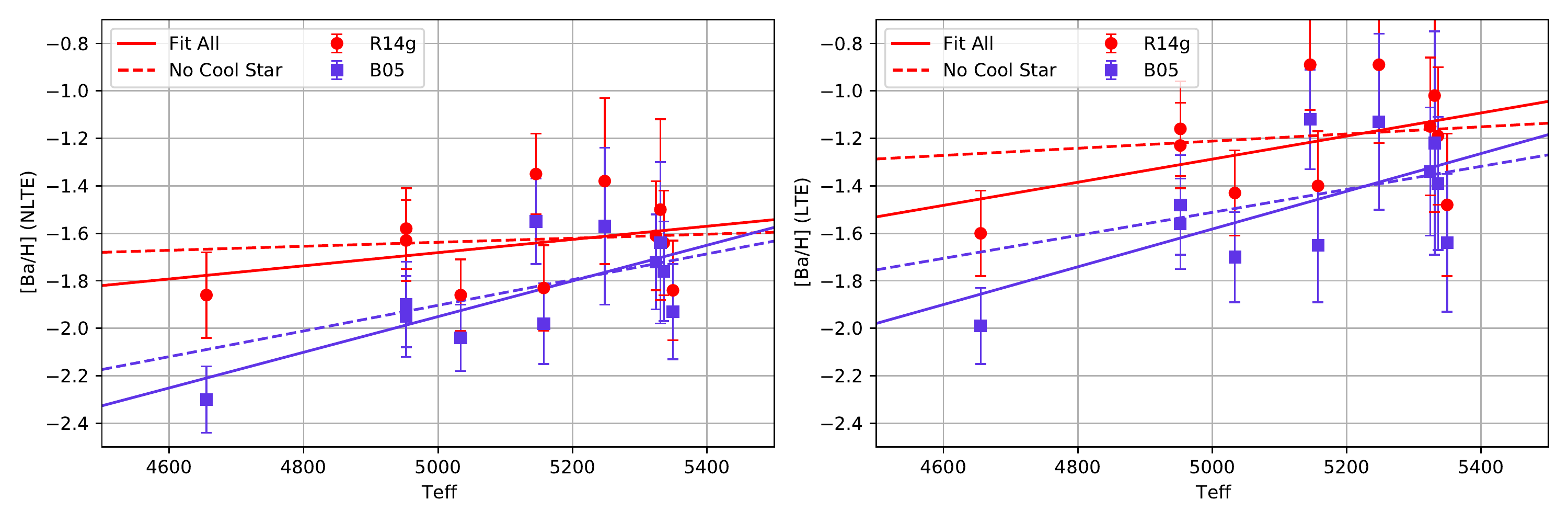}
    \caption{Comparison of [Ba/H] vs $\Teff$ slopes for different microturbulence relations. In contrast to Fig~\ref{fig:vtsys}, here we only include the high SNR clear member stars.
    The left and right panels show the NLTE and LTE [Ba/H] abundances, respectively.
    The red circles and blue squares indicate [Ba/H] derived using the R14g and B05 $\nu_t-\logg$ relations, respectively. The red and blue solid lines indicate a fit to all stars using the R14g and B05 relations, respectively; while the dashed lines indicate the fit removing the coolest star (ID 1). It is clear that the trend with $\Teff$ is flatter using the R14g relation, regardless of whether the coolest star is included.
    For the NLTE panel, the slopes of the lines (in units of dex per 100K) are 0.03 and 0.08 for R14g and B05, respectively; or 0.01 and 0.05 when removing the cold star for R14g and B05, respectively.
    }
    \label{appfig:vtrelationslope}
\end{figure}

In this paper, we have decided to pick a $\nu_t-\logg$ relation that leaves no trend in the [Ba/H] and $\Teff$ for our stars.
In Figure~\ref{fig:vtsys}, the 2nd column shows [Ba/H] vs $\Teff$, where the first three rows of Figure~\ref{fig:vtsys} have a strong systematic trend such that cooler stars have lower [Ba/H]. The bottom three rows do not have a significant trend.
The coolest stars have the highest SNR and lowest statistical uncertainty, so the different mean trend makes a significant difference on our final inferred Ba scatter.
Figure~\ref{appfig:vtrelationslope} shows this specifically for the high SNR and clear member stars for the R14g and B05 relations. The R14g $\nu_t-\logg$ relation clearly has less trend with $\Teff$ than B05.
Note that the trends are primarily driven by the two coolest and brightest stars ($\Teff \sim 4500\,$K). Because these two stars have low statistical uncertainty on their [Ba/H] abundances, the systematic effect of microturbulence can make a large difference in the inferred [Ba/H] scatter.

We have decided to adopt the quadratic $\nu_t-\logg$ relation from the giant stars in \citet{Roederer14c} (R14g) as our fiducial results. R14g have the highest SNR spectra of metal-poor giants with the largest wavelength coverage out of all these data samples.
The results of this paper would not change if we used the very similar M08 or R14 relations instead.
However, this $\nu_t-\logg$ relation is different from most previous studies of dwarf galaxy stellar abundances \citep{Ji16c,Ji18,Roederer16b}.
Thus in Appendix~\ref{app:basys}, we give all results using the $\nu_t-\logg$ relation from B05 as well, which matches those previous abundance studies.
We note that adding NLTE effects for Ba exacerbates the trend for [Ba/H] vs $\Teff$ when using the B05 $\nu_t$-$\logg$ relation, because the NLTE corrections are larger (more negative) when microturbulence is higher.
It is also worth noting that past studies of Ret~II had clear [Ba/H] trends with temperature that can be explained by microturbulence.

\subsection{Microturbulence Scatter}
Typically, a systematic uncertainty of ${\approx}$0.2 \kms is adopted for microturbulence, which accounts for the systematic mean differences described above.
However, because we are interested in the abundance scatter within Ret~II, another crucial value is the \emph{intrinsic scatter} in microturbulence around a ``true'' $\nu_t-\logg$ relation, i.e., changes in the atmospheric structure that are unmodeled by $\logg$ alone.
This error is likely \emph{smaller} than the observational scatter, because the microturbulence measurements themselves are noisy.

Examining our five datasets, two have a scatter of ${\sim}0.2$ \kms (C13, J15), while three have a scatter of ${\sim}0.1$ \kms (B05, M08, R14).
Using a smaller intrinsic $\nu_t$ scatter would increase the significance of our scatter detections, while a larger intrinsic $\nu_t$ scatter reduces the significance.
Since we have adopted the mean $\nu_t-\logg$ relation from R14g, we also decide to adopt the intrinsic scatter of $0.13\kmsec$ from that data sample.
In our systematic investigations using the B05 sample, we adopt the corresponding intrinsic scatter of $0.12\kmsec$.

\begin{deluxetable}{l|cc|cc}
    \tablecolumns{5}
    \tablecaption{\label{apptab:vtparams}Fit parameters for $\nu_t$}
    \tablehead{Sample & Quadratic Fit & $\sigma$ & Linear Fit & $\sigma$}
    \startdata
 B05        & $0.1001 \log g^2 + -0.7394 \log g + 2.847$ & 0.12 & $-0.2527 \log g + 2.316$ & 0.15 \\
 C13        & $0.1048 \log g^2 + -0.7744 \log g + 2.965$ & 0.18 & $-0.2189 \log g + 2.300$ & 0.20 \\
 J15        & $0.1307 \log g^2 + -0.9812 \log g + 3.322$ & 0.21 & $-0.5217 \log g + 2.973$ & 0.22 \\
 M08        & $0.0175 \log g^2 + -0.3242 \log g + 2.009$ & 0.08 & $-0.2545 \log g + 1.944$ & 0.08 \\
 R14        & $0.0471 \log g^2 + -0.3474 \log g + 1.969$ & 0.18 & $-0.1201 \log g + 1.764$ & 0.18 \\
 R14 giants & $0.0386 \log g^2 + -0.3313 \log g + 1.960$ & 0.13 & $-0.2247 \log g + 1.897$ & 0.10 \\    
    \enddata
\end{deluxetable}

\section{Systematic Effects on Barium Scatter}\label{app:basys}

Here we explore the effect of different data subsets and microturbulence relations on the main result of this paper, the [Ba/H] mean and scatter.
For the data samples, we consider permutations of membership (clear members only vs including candidate members) and MULTI NLTE/MARCS vs MOOG LTE/ATLAS.
For the microturbulence relations, we use the fiducial R14 giants (R14g) relation, as well as the B05 relation that has a higher microturbulence for the coolest/lowest gravity giants.
For each of these permutations, we fit the two-component Ba scatter model described in Section~\ref{sec:bascat}.
Note that while our fiducial model is run with a very large number of steps, for the other models we only sampled to reach ${\gtrsim}100$ effective samples, and thus the uncertainties and limits on the parameters will be less accurate.

Table~\ref{apptab:basys} gives the results of the model fits. The first row is our fiducial value, while the other rows show various data permutations.
$\mu_1$ and $\sigma_1$ are the most important values, indicating the mean and intrinsic spread on the detected [Ba/H] abundances. $\mu_2$ and $\sigma_2$ are the mean and scatter of the undetected [Ba/H] component, which is not well-constrained given that no low [Ba/H] abundances were detected. $p_2$ is the fraction of stars in the undetected [Ba/H] component, i.e., $1-p_2$ is the fraction of $r$-enhanced stars.
The uncertainties are $1\sigma$, and the limit on $\sigma_1$ is a 95\% limit.

\begin{figure}
    \centering
    \includegraphics[width=\linewidth]{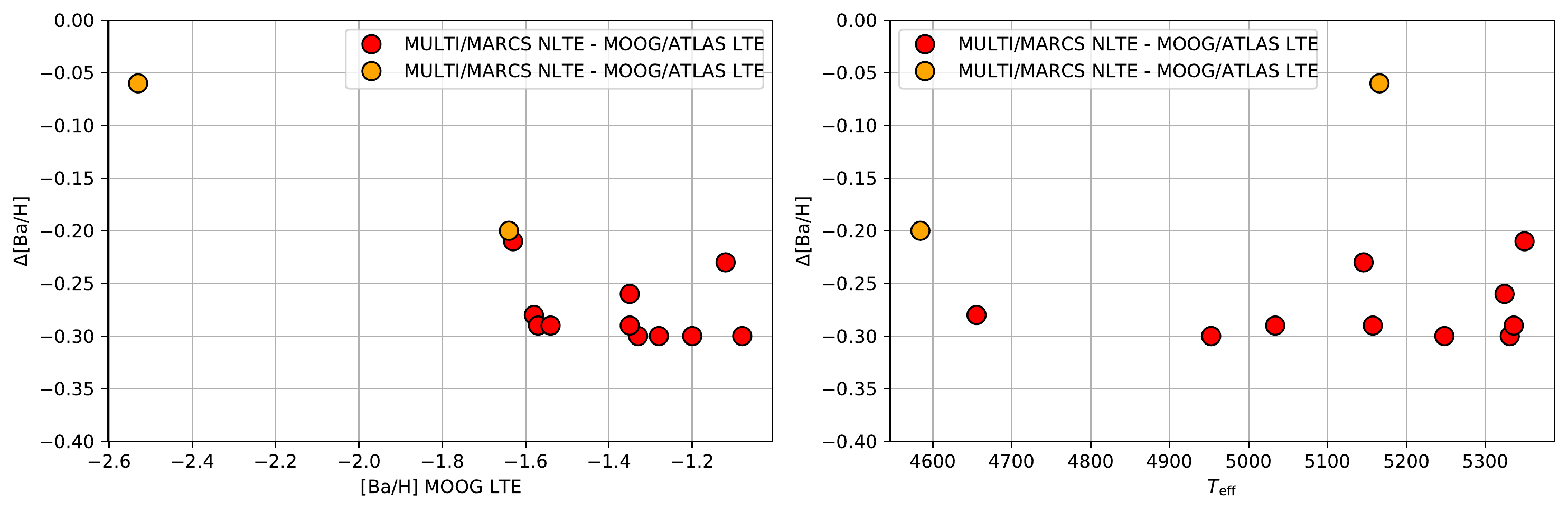}
    \caption{Differences between NLTE (MULTI/MARCS) and LTE (MOOG/ATLAS) abundances for stars with SNR $> 25$. Left: differences as a function of [Ba/H] (LTE).
    Right: differences as a function of $\Teff$. The NLTE correction for saturated lines clusters closely around $\Delta$[Ba/H]$=-0.27 \pm 0.04$ (the outlier is the candidate member star 14 with a relatively low [Ba/H] abundance).}
    \label{appfig:nlter14}
\end{figure}

We point out three main conclusions of Table~\ref{apptab:basys}.
First, the MULTI NLTE/MARCS mean [Ba/H] abundances ($\mu_1$) are typically lower than the MOOG LTE/ATLAS abundances by about 0.3 dex. Figure~\ref{appfig:nlter14} shows that the typical [Ba/H] correction going from MOOG to MULTI is $-0.27 \pm 0.04$ dex in a way that is fairly close to a constant offset. This is a result both of the different model atmospheres as well as the effect of NLTE.
Second, when considering just the clear member stars, none of the models detect a significant [Ba/H] dispersion $\sigma_1$. The constraint is stronger when using NLTE, but weaker when using the B05 microturbulence relation instead of the R14g microturbulence relation.
This is driven primarily by the coolest Ret~II star (ID 1), which is most affected by the different microturbulence relations (Figure~\ref{fig:vtsys}).
Third, when including the candidate member stars, all the upper limits on $\sigma_1$ get looser, and actually in one case (B05 MOOG with candidates) the intrinsic dispersion is resolved at $2\sigma$.
This is predominantly because of the outlier star 14, which has a weak but detected Ba line. If this star is actually part of Ret~II, then our two-component model for [Ba/H] is likely insufficient to describe the data because star 14 is well outside of the main peak of [Ba/H] detections, but well above the more stringent [Ba/H] upper limits.

\begin{figure}
    \centering
    \includegraphics[width=\linewidth]{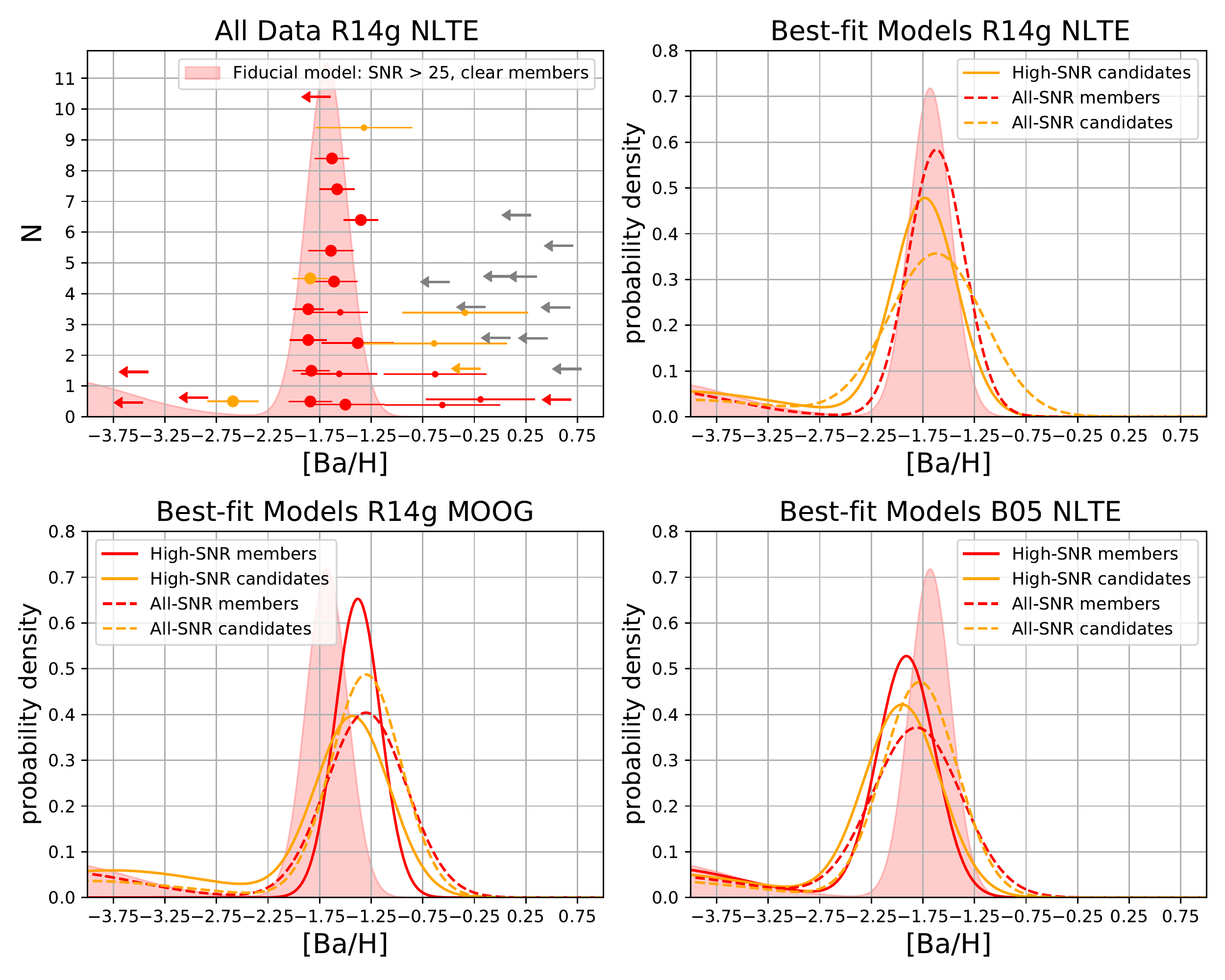}
    \caption{Exploration of different best-fit models by permuting the data sample (top-right panel),
    radiative transfer (bottom-left panel), and $\logg-\nu_t$ relation (bottom-right panel).}
    \label{appfig:bahdistrs}
\end{figure}

After examining Table~\ref{apptab:basys} and Figure~\ref{fig:vtsys}, we decided that using the R14g MULTI/NLTE results with only clear members is the most reliable measurement.
It is clear that using NLTE and definite members will result in a better measurement, and eliminating the trend with stellar parameters discussed in Appendix~\ref{app:vtba} justifies using R14g instead of B05.
However, for completeness, we show several permutations of best-fit [Ba/H] distributions in Figure~\ref{appfig:bahdistrs}. The top-left panel shows the R14g and NLTE abundances used in the main paper, but plotting all low-SNR detections as small data points and low-SNR upper limits as grey arrows. As also seen in the right column of Figure~\ref{fig:vtsys}, the low-SNR data are skewed towards higher [Ba/H] abundances primarily due to bad sky subtraction (Figure~\ref{fig:vltfit}).
The top-right panel shows three alternate fits to different permutations of data used (members and candidates; high- and low-SNR data).
The bottom left panel shows the effect of changing the radiative transfer, and the bottom-right panel shows the effect of changing the microturbulence relation. These differences make a relatively small change to the Ba dispersion (which is not resolved) but a fairly large change to the mean abundance.

\begin{deluxetable}{l|ccc|cc|c}
\tablecolumns{7}
\tabletypesize{\scriptsize}
\tablecaption{\label{apptab:basys} [Ba/H] distribution fits (stars with SNR $>25$)}
\tablehead{Data & $\mu_1$ & $\sigma_1$ & $\sigma_1$ limit & $\mu_2$ & $\sigma_2$ & $p_2$}
\startdata
R14g MULTI NLTE members         & $-1.68^{+0.07}_{-0.07}$ & $ 0.05^{+0.08}_{-0.03}$ & $<0.20$ & $-4.32^{+0.49}_{-0.46}$ & $ 0.08^{+0.32}_{-0.06}$ & $ 0.28^{+0.12}_{-0.10}$ \\
\hline
R14g MOOG LTE members           & $-1.38^{+0.06}_{-0.06}$ & $ 0.06^{+0.09}_{-0.04}$ & $<0.22$ & $-4.26^{+0.53}_{-0.50}$ & $ 0.09^{+0.32}_{-0.07}$ & $ 0.28^{+0.12}_{-0.10}$ \\
B05 MULTI NLTE members         & $-1.91^{+0.07}_{-0.06}$ & $ 0.11^{+0.10}_{-0.08}$ & $<0.28$ & $-4.31^{+0.49}_{-0.47}$ & $ 0.09^{+0.36}_{-0.07}$ & $ 0.26^{+0.12}_{-0.10}$ \\
B05 MOOG LTE members           & $-1.60^{+0.07}_{-0.07}$ & $ 0.14^{+0.09}_{-0.09}$ & $<0.30$ & $-4.22^{+0.48}_{-0.51}$ & $ 0.11^{+0.37}_{-0.09}$ & $ 0.23^{+0.14}_{-0.13}$ \\
\hline
R14g MULTI NLTE with candidates  & $-1.73^{+0.07}_{-0.08}$ & $ 0.12^{+0.11}_{-0.09}$ & $<0.31$ & $-4.12^{+0.57}_{-0.56}$ & $ 0.23^{+0.55}_{-0.20}$ & $ 0.26^{+0.12}_{-0.11}$ \\
R14g MOOG LTE with candidates    & $-1.43^{+0.07}_{-0.08}$ & $ 0.13^{+0.13}_{-0.09}$ & $<0.36$ & $-3.77^{+0.62}_{-0.69}$ & $ 0.62^{+0.28}_{-0.50}$ & $ 0.29^{+0.14}_{-0.11}$ \\
B05 MULTI NLTE with candidates  & $-1.95^{+0.07}_{-0.04}$ & $ 0.19^{+0.10}_{-0.10}$ & $<0.37$ & $-4.25^{+0.52}_{-0.51}$ & $ 0.10^{+0.46}_{-0.08}$ & $ 0.22^{+0.12}_{-0.09}$ \\
B05 MOOG LTE with candidates    & $-1.69^{+0.11}_{-0.10}$ & $ 0.24^{+0.11}_{-0.10}$ & $<0.46$ & $-3.83^{+0.74}_{-0.61}$ & $ 0.40^{+0.44}_{-0.37}$ & $ 0.25^{+0.15}_{-0.13}$ \\
\enddata
\end{deluxetable}

\section{Additional Chemical Abundances} \label{app:moreelems}
We provide a table of member stars with sufficiently high S/N in the M2FS HiRes Mg b data to measure detailed chemical abundances. This illustrates the usefulness of the M2FS Mg Wide configuration for measuring detailed chemical abundances in metal-poor stars.

\begin{deluxetable*}{rr|rr|rr|rr|rr|rr|rr}
\tablecolumns{14}
\tablecaption{\label{tab:extra_abunds}M2FS Mg b Abundances}
\tablehead{ID & SNR & [Mg/H] & $\sigma_{\rm Mg}$ & [Ca/H] & $\sigma_{\rm Ca}$ & [Ti/H] & $\sigma_{\rm Ti}$ & [Cr/H] & $\sigma_{\rm Cr}$ & [Fe/H] & $\sigma_{\rm Fe}$ & [Nd/H] & $\sigma_{\rm Nd}$ }
\startdata
  2 & 39.3 & $-2.48$ & $ 0.12$ & $-2.21$ & $ 0.07$ & $-2.29$ & $ 0.10$ & $-2.83$ & $ 0.13$ & $-2.63$ & $ 0.11$ & $-1.20$ & $ 0.09$ \\
  3 & 34.2 & $-2.59$ & $ 0.13$ & $-2.38$ & $ 0.07$ & $-2.57$ & $ 0.08$ & $-3.35$ & $ 0.11$ & $-2.78$ & $ 0.13$ & $-1.35$ & $ 0.08$ \\
  4 & 37.3 & $-2.55$ & $ 0.11$ & $-2.47$ & $ 0.08$ & $-2.71$ & $ 0.08$ & $-3.38$ & $ 0.09$ & $-3.14$ & $ 0.08$ & \nodata & \nodata \\
  5 & 19.4 & $-2.52$ & $ 0.13$ & $-1.86$ & $ 0.13$ & $-1.65$ & $ 0.19$ & $-2.10$ & $ 0.20$ & $-1.93$ & $ 0.17$ & $-0.89$ & $ 0.15$ \\
  6 & 23.5 & $-2.75$ & $ 0.13$ & $-2.24$ & $ 0.15$ & $-2.55$ & $ 0.10$ & \nodata & \nodata & $-2.81$ & $ 0.12$ & \nodata & \nodata \\
  7 & 13.8 & $-2.67$ & $ 0.22$ & \nodata & \nodata & \nodata & \nodata & \nodata & \nodata & \nodata & \nodata & \nodata & \nodata \\
  8 & 15.2 & $-2.03$ & $ 0.16$ & $-1.52$ & $ 0.17$ & $-1.76$ & $ 0.14$ & $-2.42$ & $ 0.23$ & $-2.17$ & $ 0.12$ & $-0.90$ & $ 0.12$ \\
  9 & 17.0 & $-2.72$ & $ 0.13$ & \nodata & \nodata & $-2.55$ & $ 0.12$ & $-3.32$ & $ 0.14$ & $-2.67$ & $ 0.11$ & \nodata & \nodata \\
 10 & 15.3 & $-2.09$ & $ 0.18$ & \nodata & \nodata & $-2.44$ & $ 0.17$ & $-3.49$ & $ 0.16$ & $-2.86$ & $ 0.16$ & \nodata & \nodata \\
 12 & 15.2 & $-2.84$ & $ 0.15$ & $-1.94$ & $ 0.16$ & $-2.40$ & $ 0.13$ & $-3.07$ & $ 0.22$ & \nodata & \nodata & \nodata & \nodata \\
 13 &  9.9 & $-2.60$ & $ 0.16$ & \nodata & \nodata & \nodata & \nodata & $-2.46$ & $ 0.27$ & \nodata & \nodata & \nodata & \nodata \\
\enddata
\end{deluxetable*}

\end{document}